\documentclass{aa}  

\usepackage{graphicx}
\usepackage{physics}
\usepackage{mathtools}
\usepackage{amssymb}
\usepackage{amsmath}
\usepackage{tablefootnote}
\usepackage[varg]{txfonts}
\begin{document}

   \title{The spectroastrometric detectability of nearby \\ Solar System-like exomoons}

   \author{Q. B. van Woerkom
          \inst{1}\fnmsep\inst{2}
          \and
          E. Kleisioti\inst{1}\fnmsep\inst{2}
          }

   \institute{Leiden Observatory, Leiden University, P.O. Box 9513, 2300 RA Leiden, The Netherlands \\
              \email{woerkom@mail.strw.leidenuniv.nl}
         \and
             Faculty of Aerospace Engineering, Delft University of Technology, Kluyverweg 1, 2629 HS, Delft, The Netherlands\\
             }

   \date{Received Month DD, 20YY; accepted Month DD, 20YY}

 
  \abstract
   {Though efforts to detect them have been made with a variety of methods, no technique can claim a successful, confirmed detection of a moon outside the Solar System yet. Moon detection methods are restricted in capability to detecting moons of masses beyond what formation models would suggest, or they require surface temperatures exceeding what tidal heating simulations allow.}
   {We expand upon spectroastrometry, a method that makes use of the variation of the centre of light with wavelength as the result of an unresolved companion, which has previously been shown to be capable of detecting Earth-analogue moons around nearby exo-Jupiters, with the aim to place bounds on the types of moons detectable using this method.}
   {We derived a general, analytic expression for the spectroastrometric signal of a moon in any closed Keplerian orbit, as well as a new set of estimates on the noise due to photon noise, pointing inaccuracies, background and instrument noise, and a pixelated detector. This framework was consequently used to derive bounds on the temperature required for Solar System-like moons to be observable around super-Jupiters in nearby systems, with $\epsilon$ Indi Ab as an archetype.}
   {We show that such a detection is possible with the ELT for Solar System-like moons of moderate temperatures (150-300 K) in line with existing literature on tidal heating, and that the detection of large (Mars-sized or greater) icy moons of temperatures such as those observed in our Solar System in the very nearest systems may be feasible.}
   {}

   \keywords{Astrometry - detection - tidal interaction - natural satellites (extrasolar)}

    \maketitle
%

\section{Introduction} \label{sec:intro}
The Solar System hosts a variety of large moons, presenting environments as diverse and scientifically interesting as its planets. As the count of exoplanets found so far is in the thousands, it is natural to ask whether these extrasolar planets hide an equal number of extrasolar moons. Secondary or population-level effects of such moons have already been found (e.g. \citealt{Kenworthy2015ModelingExomoons, Hippke2015ONPEAK, Teachey2018HEK.I, Oza2019SodiumExoplanets, Saillenfest2023ObliqueRadii}), and so the question is not whether these moons exist, but rather what they look like, and in what systems and around which planets can they be found.

The detection and characterisation of such satellites accompanying extrasolar planets holds the potential to further our understanding of planet formation, evolution, and habitability. Different moon and planet formation scenarios and their outcomes have been linked to predictions of or requirements on, for example, the time of formation of the satellites \citep{Cilibrasi2018SatellitesMoons}, their size and mass relative to their host \citep{Nakajima2022LargeMoons, Canup2006APlanets, Hansen2019FormationCapture}, circumplanetary disc (CPD) composition \citep{Batygin2020FormationSatellites, Oberg2023CircumplanetaryComposition}, host magnetosphere \citep{Canup2006APlanets}, host mass \citep{Oberg2023CircumplanetaryComposition}, instellation or host migration history \citep{Heller2015ConditionsStars, Heller2015WATERPLANETS}, and orbital properties \citep{Li2020CaptureNereid}. While the Solar System is host to a large number of moons, the limited scenario we are presented with cannot conclusively bear witness to any of these analyses: validation of these formation studies would require the detection and characterisation of satellites of planets outside the Solar System. 

Moreover, moons with the mass of Mars or greater (which are not found in the Solar System) are promising targets for habitability studies \citep{Lammer2014, Williams1997HabitablePlanets, Dobos2022AExomoons, Forgan2016ExomoonHeating}, and the observationally confirmed existence of subsurface oceans on Titan \citep{Beuthe2015TidalCo., Bills2011RotationalTitan, Baland2011TitansOcean}, Enceladus \citep{Beuthe2016CrustalMoons}, among others, and the likely current or past existence of such oceans on Triton \citep{Nimmo2015PoweringGeology, Schenk2021Triton:Charon, McKinnon2014Triton, Gaeman2012SustainabilityInterior} and other Solar System bodies (e.g. \citealt{Hussmann2006SubsurfaceObjects, Burnett2023Spin-OrbitInterior, Rovira-Navarro2023Thin-shellWorlds, Nimmo2016ReorientationPluto, Bagheri2022TheSystem, Bierson2022ASatellites, Nimmo2016OceanSystem}) hold promise for the potential habitability of such worlds, even at sizes observed in the Solar System. This makes Solar System-like satellites objects of interest concerning habitability in their own right, not just as extensions of the population of small icy and rocky planets orbiting stars directly.

Current searches for exomoons are limited primarily to those using transit timing variations (TTVs) and transit duration variations (TDVs), which can yield orbital and mass information on the moon \citep{Kipping2009TransitExomoon, Kipping2009TransitII, Heller2014FormationMoons, Teachey2018HEK.I, Teachey2020LooseKepler-1625b, Fox2021ExomoonExomoons, Kipping2022AData, Kipping2022AnB-i}. For this method, unfortunately, the effects of even large moons are degenerate with those produced by other, potentially unobserved planets in the system \citep{Fox2021ExomoonExomoons, Kipping2020IMPOSSIBLEEXOMOON}. Consequently, candidate moons are met with skepticism \citep{Teachey2020LooseKepler-1625b, Fox2021ExomoonExomoons, Kipping2020AnCandidates, Tokadjian2022ProbingB}, especially as they have thus far required invoking eccentric formation scenarios \citep{Hamers2018I, Hansen2019FormationCapture, Kipping2022AnB-i}. Other detection and characterisation methods have been proposed, from those directly analogous to those used in planet detection such as radial velocity measurements of the host planet \citep{Ruffio2023DetectingProspects, Vanderburg2018DetectingExoplanets} or transit spectroscopy \citep{Kaltenegger2010CHARACTERIZINGEXOMOONS, Limbach2021OnObjects} to the detection of thermal excesses in direct imaging data \citep{Limbach2013OnExomoons, Kleisioti2021CouldB, Kleisioti2023TidallyCharacterization} or microlensing \citep{Han2002ONMICROLENSING, Han2008MICROLENSINGEXOPLANETS, Hwang2018OGLE-2015-BLG-1459L:Microlensing}. These also face major limitations, however. The first two methods, with current instrumentation, can unfortunately only detect binary-like satellites in terms of size and mass \citep{Lazzoni2022DetectabilityDwarfs}, whereas the third requires exceptional tidal heating rates. Microlensing detections are isolated events, and therefore are difficult to confirm and cannot be followed up on.

In this analysis, we build upon previous work by \citet{Cabrera2007DetectingEvents}, who discussed the possibility of detecting extrasolar satellites by astrometry of the planet or photometric phenomena induced by mutually eclipsing and dimming events, and \citet{Agol2015THEEXOMOONS}, who put these two together and discussed the possibility of detecting extrasolar satellites by use of a method called spectroastrometry, in which one measures the differential on-sky position between the light originating from a system in different filters. A shift in the centre of light between two filters cannot be the result of a point-source or symmetric object (e.g. rings), and so explanation of such a shift in a planet signal requires the presence of a (potentially unresolved) second object. This method has previously found use in characterisation of close-separation stellar binaries and active galactic nuclei at scales well below the diffraction limit, as the centroid can be measured to precision better than the diffraction or seeing limits \citep{Bailey1998DetectionSpectro-astrometry, Bailey1998Spectroastrometry:Scales, Porter2004OnFluxes, Whelan2008Spectro-astrometry:Applications}.

\citet{Bailey1998Spectroastrometry:Scales} already predicted that `... a large telescope using low-order adaptive optics to achieve image sizes of $\sim0.1$ arc second in the IR should be able to use spectroastrometry to make measurements to $\sim100$ micro-arcsec', which is roughly equivalent to the separations of Solar System-like moons at distances of the order of tens of parsecs. We derive improved estimates of the signal and noise for spectroastrometry of photometric points, and we show that the upcoming class of extremely large telescopes (ELTs), with the European ELT as archetype, will indeed be capable of detecting and characterising nearby mild-to-warm tidally heated exomoons (THEMs) of sizes and separations like those found in our Solar System and compatible with moon formation theory, with relatively little regard for inclination or orientation of the satellite orbit.

In contrast, satellites with the expected mass ratios of $\sim 10^{-4}$ compared to their host predicted by \citet{Canup2006APlanets} at medium-to-wide planet-moon separations will remain elusive even with future instrumentation for most other methods \citep{Ruffio2023DetectingProspects, Lazzoni2022DetectabilityDwarfs}. Additionally, those other methods impose a restrictive orientation of the orbit of the satellite (for planet-transiting moons or radial velocity detections; \citealt{Lazzoni2022DetectabilityDwarfs}) or require a fortituous transiting orientation of the host planet with respect to its star \citep{Kipping2009TransitExomoon, Kipping2009TransitII}. We show that spectroastrometry only imposes a weak orientation preference on observations that is complementary to that required for the radial velocity and planet-transit methods.

We structure the analysis as follows: we begin by presenting the spectroastrometric signal, noise, and some their basic and derived properties in Secs.~\ref{subsec:spectroastrometric_signal} to \ref{subsec:min_moon_flux}, followed by a description of our benchmark scenario in Sec.~\ref{subsec:benchmark_system}. The results for this benchmark scenario are presented in Sec.~\ref{sec:results}, and the limitations of our analysis, the repercussions with respect to current-generation telescopes, and a comparison with other moon detection and characterisation methods are presented in Sec.~\ref{sec:discussion}. We conclude matters in Sec.~\ref{sec:conclusion}.

\section{Methods} \label{sec:methods}
\begin{figure}
    \centering
    \includegraphics[width=.5\textwidth]{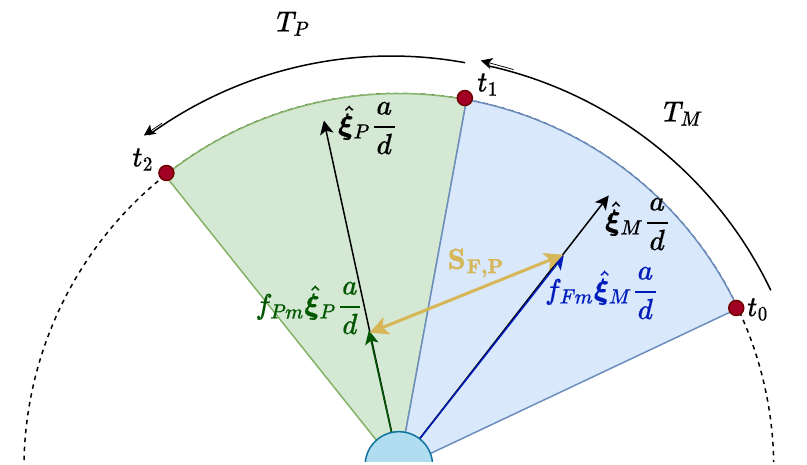}
    \caption{Illustration of the geometry underlying the calculation of the spectroastrometric signal in the face-on case ($p=1$).}
    \label{fig:orbit_explanation}
\end{figure}
In the following, we present the framework upon which our analysis is based. It is structured as follows: in Secs.~\ref{subsec:spectroastrometric_signal} and \ref{subsec:kepler_orbit_signal} we introduce the spectroastrometric signal and some of its basic properties, followed by a discussion on the contaminating noise sources in Sec.~\ref{subsec:noise_sources} and the consequences on observation design in Sec.~\ref{subsec:obs_time}. We then derive an expression for the minimum flux a moon must have to be detectable at a given signal-to-noise ratio given any instrument specifications in Sec.~\ref{subsec:min_moon_flux}, and introduce the model scenario used to evaluate the efficacy of spectroastrometry in detecting Solar System-like moons in Sec.~\ref{subsec:benchmark_system}.

\subsection{The centroid and the spectroastrometric signal} \label{subsec:spectroastrometric_signal}
\noindent In defining the spectroastrometric signal, we take an approach that differs from that taken by \citet{Agol2015THEEXOMOONS}; this framework allows us to compute several additional error estimates. Motivated by the idealised mathematical limit of the photon-count averaged centre of light on a detector D for ever smaller pixels, we shall define the centroid for an observation over a time interval $T_F$ through a filter $F$, $\mathbf{c}_F$, as the photon count-weighted integral over all positions $\mathbf{c}$  over the course of an observation $O$, during which a total of $N_F$ photons are counted as follows:
\begin{equation}
    \label{eq:deffiltercentroid}
    \mathbf{c}_F = \frac{1}{N_F}\int_{O,D} \mathbf{c} \dd{N_F(\mathbf{c})}.
\end{equation}
The differential photon count hitting the detector for each position $\mathbf{c}$, $\dd{N_F(\mathbf{c})}$, can be rewritten, accounting for the fact that the incoming photon rate is a function of the detector location it hits and of time, $t$. We choose to ignore the effect of the photon path through the instrument, such that we can then write $\dd{N_F(\mathbf{c})} = I_F(\mathbf{c},t)\dd{\Omega}\dd{t}\dd{S}$ with $I_F$ the photon intensity through the filter $F$ originating from the point $\mathbf{c}$ on the celestial sphere at time $t$, $\dd{\Omega}$ the infinitesimal area on the detector that the photons hit (expressed as solid angle on the celestial sphere), and $\dd{S}$ the infinitesimal area of the telescope aperture that the photon field passes through. If we furthermore assume that the total photon flux originating from the part of the sky we observe through the filter $F$, $F_F=\int_D I_F(\mathbf{c}, t) \dd{\Omega}$, is constant throughout the observation such that $N_F = F_F T_F S$, we obtain:
\begin{align}
    \mathbf{c}_F 
    &= \frac{1}{T} \int_T \left(\frac{1}{F_F}\int_D \mathbf{c} I_F(\mathbf{c},t)\dd{\Omega}\right)\dd{t}
\end{align}
where $\dd{S}$ was taken outside the integral as we had assumed the photon flux on the detector to be independent of the photon path through the instrument. The outermost integral is ostensibly a time-average, while 
\begin{equation}
    \mathbf{c}_F(t) = \frac{1}{F_F}\int_D \mathbf{c} I_F(\mathbf{c},t)\dd{\Omega}
\end{equation}
is the photon flux-averaged centre of light at time $t$. We have thus shown that the centroid as measured from an observation is the time-averaged value of the instantaneous centroid $\mathbf{c}_F(t)$ over the duration of that observation: $\mathbf{c}_F=\left<\mathbf{c}_F(t)\right>_T$.

We note that for an $I_F(\mathbf{c}, t)$ that is point-symmetric about some point $\mathbf{c}_0(t)$ we have $\mathbf{c}_F(t) = \mathbf{c}_0(t)$: motivated by this observation, we shall now assume that we are observing a planet-moon system (we note, however, that computation of the centroid of an observation according to Eq.~\ref{eq:deffiltercentroid} does not require this assumption), such that we can decompose the intensity originating from any given on-sky location into that originating from the planet (denoted by the subscript $p$) and that originating from the moon (with the corresponding subscript $m$). Then, by linearity of the integral:
\begin{align}
    \mathbf{c}_F 
    &= \frac{F_{Fp}}{F_F} \left<\mathbf{c}_{Fp}(t)\right>_{T_F} + \frac{F_{Fm}}{F_F} \left<\mathbf{c}_{Fm}(t)\right>_{T_F}
\end{align}
where $F_{Fp}$ (respectively $F_{Fm}$) is the flux in the filter $F$ due to the planet (respectively moon). We observe that so long as the point-spread function (PSF) for our telescope is point-symmetric, we have that the components $I_{Fp}(\mathbf{c}, t)$ and $I_{Fm}(\mathbf{c}, t)$ of our intensity are respectively point-symmetric about the on-sky position of the planet and moon, which we shall denote $\mathbf{c}_p$ and $\mathbf{c}_m$, such that finally:
\begin{align}
    \mathbf{c}_F 
    &= \left<\mathbf{c}_{p}(t)\right>_{T_F} + \frac{F_{Fm}}{F_F} \left<\mathbf{c}_{m}(t) - \mathbf{c}_{p}(t)\right>_{T_F}, \label{eq:centroid_location_timeavg}
\end{align}
analogous to Eq.~4 in \citet{Agol2015THEEXOMOONS}. We must note, though, that here we have shown that filter fluxes must be expressed in photon fluxes, not energy fluxes, for this relation to hold. For notational brevity, we shall henceforth write $\mathbf{c}_{m}(t) - \mathbf{c}_{p}(t) = \mathbf{c}_{mp}(t)$ for the on-sky projected angular separation of the two bodies. In imitation of the approach \citet{Agol2015THEEXOMOONS} take to forego the necessity of a reference position, we then define the spectroastrometric signal $S_{M,P}$ as the absolute difference between the centroid location in a filter $M$ in which we expect to observe the moon (the `moon filter'; not to be confused with the notation for absolute magnitude used in other literature) and another filter $P$ in which we expect the planet to be dominant (the `planet filter'); we remark that we also differentiate between the periods over which the two observations were made, $T_M$ and $T_P$ respectively:
\begin{align}
    S_{M,P} &= \abs{\mathbf{c}_M - \mathbf{c}_P} \nonumber \\
     \label{eq:spectroastrometric_signal} &= \abs{\frac{F_{Mm}}{F_M} \left<\mathbf{c}_{mp}(t)\right>_{T_M} - \frac{F_{Pm}}{F_P} \left<\mathbf{c}_{mp}(t)\right>_{T_P}}
\end{align}
where we have assumed that the time-averaged position of the planet over both the periods $T_M$ and $T_P$ is identical. As we shall encounter these quantities more often, let us denote the fraction of the flux in the band $M$ (respectively $P$) due to the moon, $\frac{F_{Mm}}{F_M}$ (respectively $\frac{F_{Pm}}{F_P}$), as  $f_{Mm}$ (respectively $f_{Pm}$).

\subsection{The spectroastrometric signal for closed Keplerian orbits}
\label{subsec:kepler_orbit_signal}
For $\left<\mathbf{c}_{mp}(t)\right>$ an analytic solution exists in terms of known quantities from orbital mechanics for all closed Keplerian orbits; its derivation is described in Sec.~\ref{sec:angular_separation_kepler}. One can then show that the expected value of the signal for a given observation of a moon with unconstrained inclination $i$ and $\omega$ (i.e. a flat prior on both $i$ and $\omega$) is given by:
\begin{align}
    S_{M,P} =  \label{eq:calculated_signal}\frac{\gamma af_{Mm}}{d}\abs{\hat{\boldsymbol{\xi}}_F - \tfrac{f_{Pm}}{f_{Mm}}\hat{\boldsymbol{\xi}}_P}
\end{align}
with $\gamma$ a parameter describing our knowledge of the inclination $i$ and orientation (through $\omega$) of the moon; if $i$ and $\omega$ are unconstrained (i.e. uniformly distributed), we have $p\approx0.842$ if instead the moon is positioned in the worst possible orientation (edge-on i.e. $i=\pi/2$), we have $p=2/\pi$; in the best possible orientation (face-on i.e. $i=0$) we have $p=1$. $\hat{\boldsymbol{\xi}}_M$ (respectively $\hat{\boldsymbol{\xi}}_P$) is the non-dimensionalised time-averaged in-orbit position over the time period $T_M$ (respectively $T_P$) of the moon, and $d$ is the system-observer distance. A derivation of Eq.~\ref{eq:calculated_signal} and the values of the quantity $p$ as well as a more extensive derivation of $\hat{\boldsymbol{\xi}}_M$ (and $\hat{\boldsymbol{\xi}}_P$) are provided in Sec.~\ref{sec:signal_kepler_derivation}; an illustration of the geometry involved is given in Fig.~\ref{fig:orbit_explanation}.

It will suffice for now to know that it can be shown that the time-averaged in-orbit position depends on the orbital properties of the moon according to Eq.~\ref{eq:xi} (for a more detailed discussion of this result, the reader is referred to App.~\ref{sec:angular_separation_kepler} for the derivation and App.~\ref{sec:app_properties} for a discussion on its properties):
\begin{align}
\label{eq:xi}
    \hat{\boldsymbol{\xi}} &= \frac{P}{2\pi T}
    \begin{pmatrix}
    (1-\frac{e^2}{2})\sin{E} - \frac{e}{4}\sin{2E} \\
    (1-e^2)^{1/2} \left(\frac{e}{4}\cos{2E} - \cos{E}\right)
    \end{pmatrix}\bigg\rvert_{E_0}^{E_1} -  
    \begin{pmatrix}
    \frac{3e}{2} \\
    0
    \end{pmatrix}
\end{align}
where $P$ (not to be confused with the filter $P$, which only appears in subscripts) and $e$ are the period and eccentricity of the orbit of the moon, $T$ is the timespan over which we observe in the filter of interest, and $E_0$ and $E_1$ are the eccentric anomaly at the start and end of the observation in that same filter, respectively. For a given time since periapsis, $E_0$ and $E_1$ can be calculated by solving Kepler's equation, for which a variety of solution methods exist; we employ the non-iterative method described by \citet{Markley1995KeplerSolver}.

\subsection{Sources of spectroastrometric noise}
\label{subsec:noise_sources}
There are, however, several sources of noise that will contaminate the signal. In the following, we shall describe relevant noise sources and their corresponding expressions, as follows: (1) photon shot noise, (2) pixel noise, (3) background and instrument flux noise and (4) pointing noise. We (will) note that each of these noise sources acts upon each of the measured centroids individually (that is, per filter) and independently. We will therefore consider in the following the noise that each of these sources imparts upon the measured centroid in a filter, noting that these noise sources are present in both the filter $M$ and the filter $P$. We can then combine the two into the resulting total noise for the spectroastrometric signal in Sec.~\ref{subsubsec:total_noise}.

\subsubsection{Photon noise}
The PSF of the telescope is not just the reason by which we cannot resolve the moon and planet directly on the exposure; the resulting spread in photon arrival locations causes an inherent and unavoidable variance in the measured centroid, given that we can only ever take a finite sample of the PSF. In Sec.~\ref{subsec:photon_noise_app} it is shown that under the condition of point symmetry of the PSF we have for the photon noise $\sigma_{PN}$:
\begin{equation}
\label{eq:photon_noise}
    \sigma_{PN} = \frac{\sigma_{PSF}}{\sqrt{N}}
\end{equation}
where $\sigma_{PSF}$ is the standard deviation of the PSF for a single sample (i.e. a single photon) and $N$ is the number of photons observed in the filter. For a diffraction-limited telescope, the Gaussian approximation of the Airy disc used by \citet{Agol2015THEEXOMOONS}, where $\sigma_{PSF}=0.45\lambda_c/D$ (with $\lambda_c$ the central wavelength of the observation filter and $D$ the telescope diameter), gives a useful general expression, but where available an estimate from PSF shape models would of course be preferred. As we use the ELT as archetype for the full class of extremely large telescopes (for which PSF models are not yet available), we will use the expression by \citet{Agol2015THEEXOMOONS} in our simulations, but prefer to retain $\sigma_{PSF}$ in expressions.

\subsubsection{Pixel noise}
Another source of noise derives from the fact that we are not probing the actual on-sky intensity distribution, but a pixelated version thereof. The finite size of the pixels on our detectors means that we lose some information about each detected photon. An upper bound for the noise that this introduces can be derived (see Sec.~\ref{subsec:pixel_noise_app}) to be given by
\begin{equation}
\label{eq:pixel_noise}
    \sigma_{px} = \frac{\alpha}{\sqrt{2N}}
\end{equation}
with $\alpha$ the pixel width, and $N$ again the total number of photons observed in the filter.

\subsubsection{Background and instrument flux noise}
The noise in background and instrument flux will also give rise to noise in the measured centroid. In Sec.~\ref{subsec:flux_noise_app} we show that an upper bound for this contribution is given by
\begin{equation}
\label{eq:flux_noise}
    \sigma_n = \sqrt{\frac{6}{5}}\left(\frac{1}{6}\left\lceil\frac{6\sigma_{PSF}}{\alpha}\right\rceil\right)\frac{\alpha}{\sqrt{N}}
\end{equation}
under the assumptions that the planet detection in the flux is $\geq5\sigma$, that we sample the centroid from the detector region encompassing the $3\sigma$-region of the PSF (which contains $>97\%$ of the incoming flux) and that the noise is uncorrelated between pixels.

\subsubsection{Pointing noise}
\label{subsubsec:pointing_noise}
The imperfect pointing accuracy of the telescope will also impact the extracted centroid; we can account for this by assuming that each observed photon has a further random offset dictated by the pointing accuracy of the telescope. In that case, the mathematical framework works out precisely as for the PSF, so long as the timescale on which the telescope pointing is affected by inaccuracies is significantly lower than the total time of the exposure, such that photons can be assumed to be independently affected. In the case of the ELT, the uncompensatable random errors vary on subsecond timescales \citep{Rodeghiero2021PerformanceAstrometry}, such that we deem this a reasonable assumption. METIS will achieve fine-guiding accuracies below $0.02\lambda/D$ \citep{Brandl2021METIS:Spectrograph}, which translates to a worst-case pointing offset below $\sigma_{PO}\approx1$ mas, such that we can expect the pointing inaccuracy $\sigma_p$ to be at the very greatest
\begin{equation}
    \sigma_p = \frac{\sigma_{PO}}{\sqrt{N}}.
\end{equation}
This term is negligible for the ELT, but may be important or even limiting for space telescopes in particular.

This treatment then accounts for random pointing noise; systematic pointing offsets can be removed by referencing against the reference objects used. If possible with the given coronagraphic system, the central star or another bright co-moving object would be a suitable candidate. Otherwise, a background object can be used. As such objects are in general far brighter than the planet, the corresponding spectroastrometric noise is likely to be negligible.  If the telescope reacquisition pointing accuracy between different filters allows for it, referencing against other objects may even become unnecessary. This may introduce systematic bias if there is an unknown offset between filters (e.g. due to imperfect adaptive optics), however, and so this should be carefully accounted for.

\subsubsection{Total noise}
\label{subsubsec:total_noise}
The fact that all noise sources go as $N^{-1/2}$ motivates us to combine them into one term, such that we can describe the total noise in a filter as $\sigma_{tot}=\sigma N^{-1/2}$. We can consider the noise $\sigma$ to be an `effective noise per photon', and under the assumption that all noise sources are independent we can calculate $\sigma$ to be
\begin{align}
\label{eq:sigma_filter}
    \sigma &= \sqrt{N}\sigma_{tot}=\sqrt{\sigma_{PN}^2 + \sigma_{px}^2 + \sigma_n^2 + \sigma_p^2} \nonumber \\
    &= \sqrt{\sigma_{PSF}^2 + \sigma_{PO}^2 + \left(\frac{1}{2} + \frac{1}{30}\left\lceil \frac{6\sigma_{PSF}}{\alpha}\right\rceil^2\right)\alpha^2}.
\end{align}
It is instructive to remark that $\sigma$ for the both filters, $\sigma_M$ and $\sigma_P$, is a property inherent to a given filter on a given telescope, but it does not depend on the object to be observed. The combined noise $\sigma$ affects the measured centroids of both of the filters in question. Assuming that these two measured centroids are statistically independent, the total noise in the measured spectroastrometric signal, $\sigma_S$, is then:
\begin{align}
\label{eq:total_sigma}
    \sigma_S &= \sqrt{\sigma_{tot,M}^2 + \sigma_{tot,P}^2} = \sqrt{\frac{\sigma_M^2}{N_M} + \frac{\sigma_P^2}{N_P}} \nonumber \\
    &= \frac{1}{\sqrt{S_{eff}\varepsilon}}\sqrt{\frac{\sigma_M^2(1-f_{Mm})}{F_{Mp}T_M} + \frac{\sigma_P^2(1-f_{Pm})}{F_{Pp}T_P}}
\end{align}
with $\varepsilon$ being the achromatic efficiency of the telescope and $S_{eff}$ its effective surface area, where we have used that $N_M = S_{eff}\varepsilon T_M F_{Mp}/(1-f_{Mm})$ (an analogous expression of course holds for $N_P$). 

Note that it follows directly from Eq.~\ref{eq:total_sigma} that the best planet filter $P$ to observe in so as to minimise the noise is the one in which $F_{Pp}/\sigma_P^2$ is maximised; the same conclusion holds for the moon filter $F$, so long as the moon is expected to be sufficiently luminous in this filter, too.

\subsection{Consequences for the observation time allocation}
\label{subsec:obs_time}
From Sec.~\ref{subsec:noise_sources} we can conclude an important guideline for observations: namely, one can show that given a fixed total observation time $T = T_M + T_P$, there exists an allocation between $T_M$ and $T_P$ such that $\sigma_S$ as given by Eq.~\ref{eq:total_sigma} is optimised. This time allocation can be shown to satisfy the relation
\begin{equation}
    \frac{T_M}{T_P} = \frac{\sigma_M}{\sigma_P}\sqrt{\frac{F_{Pp}}{F_{Mp}}}\sqrt{\frac{1-f_{Mm}}{1-f_{Pm}}}\approx \frac{\sigma_M}{\sigma_P}\sqrt{\frac{F_{Pp}}{F_{Mp}}}.
\end{equation}
The approximation is justified by noting that for a suitable planet filter $P$ we ought to expect that $f_{Pm}\approx~0$, and if the moon does not fully dominate in $M$, $\sqrt{1-f_{Mm}}\approx 1$ is also reasonable. As this approximated optimal time allocation requires no a priori knowledge of any tentative moon, we find it most representative to perform our evaluation of the method using this time allocation. We wish to note that this optimal time allocation is equally valid for the noise estimate used by \citet{Agol2015THEEXOMOONS}, though they do not appear to have taken note of it.

\subsection{Minimum required moon flux} \label{subsec:min_moon_flux}\
In general, unfortunately, we have no a priori knowledge of the expected flux of moons. Moreover, we are interested not in the signal-to-noise ratio produced by a specific moon, but rather we would like to answer the converse question: the luminosity a moon ought to have such that it is detectable with at least a given signal-to-noise ratio $S/N$. One can take Eqs.~\ref{eq:calculated_signal} and \ref{eq:total_sigma} to arrive, after some manipulation, at a criterion for detectability:
\begin{align}
\label{eq:detectability_criterion}
    \hat{\boldsymbol{\xi}}_M^2f_{Mm}^{2} + \left(C_M - 2 f_{Pm}\hat{\boldsymbol{\xi}}_M \cdot \hat{\boldsymbol{\xi}}_P\right)f_{Mm} \geq \nonumber \\ C_M + C_P -f_{Pm}^{2}\hat{\boldsymbol{\xi}}_P^2,
\end{align}
where
\begin{align}
    &C_{M} = \left(\frac{S/Nd\sigma_M}{a\gamma}\right)^2\frac{1}{\varepsilon S_{eff}F_{Mp}T_M}, \\ &C_{P} =  \left(\frac{S/Nd\sigma_P}{a\gamma}\right)^2\frac{1 - f_{Pm}}{\varepsilon S_{eff}F_{Pp}T_P}.
\end{align}
This is a quadratic form in $f_{Fm}$, whence one can derive the following minimum flux requirement for detectability at the specified signal-to-noise ratio:
\begin{equation}
\label{eq:min_flux_constraint}
    \frac{F_{Mm}}{F_{Mp}} \geq \frac{A + \sqrt{B}}{D}
\end{equation}
with
\begin{align}
    A &= \frac{C_M}{2} + C_P + f_{Pm}\hat{\boldsymbol{\xi}}_M \cdot \hat{\boldsymbol{\xi}}_P-f_{Pm}^{2}\hat{\boldsymbol{\xi}}_P^2 \\
    B &= \left(\frac{C_M}{2} - f_{Pm}\hat{\boldsymbol{\xi}}_M \cdot \hat{\boldsymbol{\xi}}_P\right)^2 - \hat{\boldsymbol{\xi}}_M^2\left(f_{Pm}^{2}\hat{\boldsymbol{\xi}}_P^2 - C_M - C_P\right) \\
    D &= \left(\hat{\boldsymbol{\xi}}_M - f_{Pm}\hat{\boldsymbol{\xi}}_P\right)^2 - C_P.
\end{align}
This, should be stressed, holds for all closed Keplerian orbits, given the expression for $\hat{\boldsymbol{\xi}}$ in the form of Eq.~\ref{eq:xi}, so long as $D>0$. This latter constraint follows when writing Eq.~\ref{eq:detectability_criterion} as a form that is quadratic in $F_{Mm}/F_{Mp}$, and it is satisfied roughly when $C_P\lesssim 1$ such that
\begin{equation}
    \frac{S/N \sigma_P}{\sqrt{\epsilon S_{eff}F_{Pp} T_P}} \lesssim \frac{a\gamma}{d}.
\end{equation}

Note that the moon fluxes in either band are, of course, both unknown quantities; as $f_{Pm}$ is relatively low (close to zero) if a suitable planet filter $P$ was chosen, it is more informative (and receptive to implicit computation) to put constraints on $F_{Mm}$ as a function of $f_{Pm}$ than the other way around. If one can relate $f_{Mm}$ and $f_{Pm}$ to a set of defining parameters (say, for example, the surface temperature and radius of a moon), it is then possible to solve the constraint posed by Eq.~\ref{eq:min_flux_constraint} implicitly, given a set of orbital and observation parameters.

\subsection{Model scenario} \label{subsec:benchmark_system}
\begin{table}[]
\caption{\label{tab:benchmark_system} Telescope and system model parameters used in the benchmark scenario.}
\centering
\begin{tabular}{lll}
\hline
\textbf{Quantity}            & \textbf{Value} & \textbf{Reference}        \\ \hline
\multicolumn{3}{c}{\textbf{METIS/ELT properties}}                         \\ \hline
Planet filter $P$            & $M'$           & -                         \\
Moon filter $M$              & $N2$           & -                         \\
Diameter [m]                 & 37             & C20                \\
System throughput [-]        & 0.36           & C20                \\
Atmospheric transmission [-] & 0.8            & J13, N12       \\
$\sigma_{PSF}$ in $P$ [mas]  & 12.0          & -               \\
$\sigma_{PSF}$ in $M$ [mas]  & 28.1          & -               \\ \hline
\multicolumn{3}{c}{\textbf{Model system properties}}                      \\ \hline
Reference distance [pc]      & 3.6481         & P16, V23 \\
Planet mass [M$_{\textrm{Jup}}$]          & 3.25           & F19                  \\
Age [Gyr]                    & $>3.0$            & F19        \\ \hline                
\end{tabular}
\tablebib{C20: \citet{Carlomagno2020METISPerformance}; J13: \citet{Jones2013AnParanal}; N12: \citet{Noll2012AnRange}; P16: \citet{Prusti2016TheMission}; V23: \citet{Vallenari2023Gaia3}; F19: \citet{Feng2019DetectionData}. \citet{Brandl2021METIS:Spectrograph} give an overview of the METIS optical system and pixel sizes.}
\end{table}

\begin{figure}
    \centering
    \includegraphics[width=.5\textwidth]{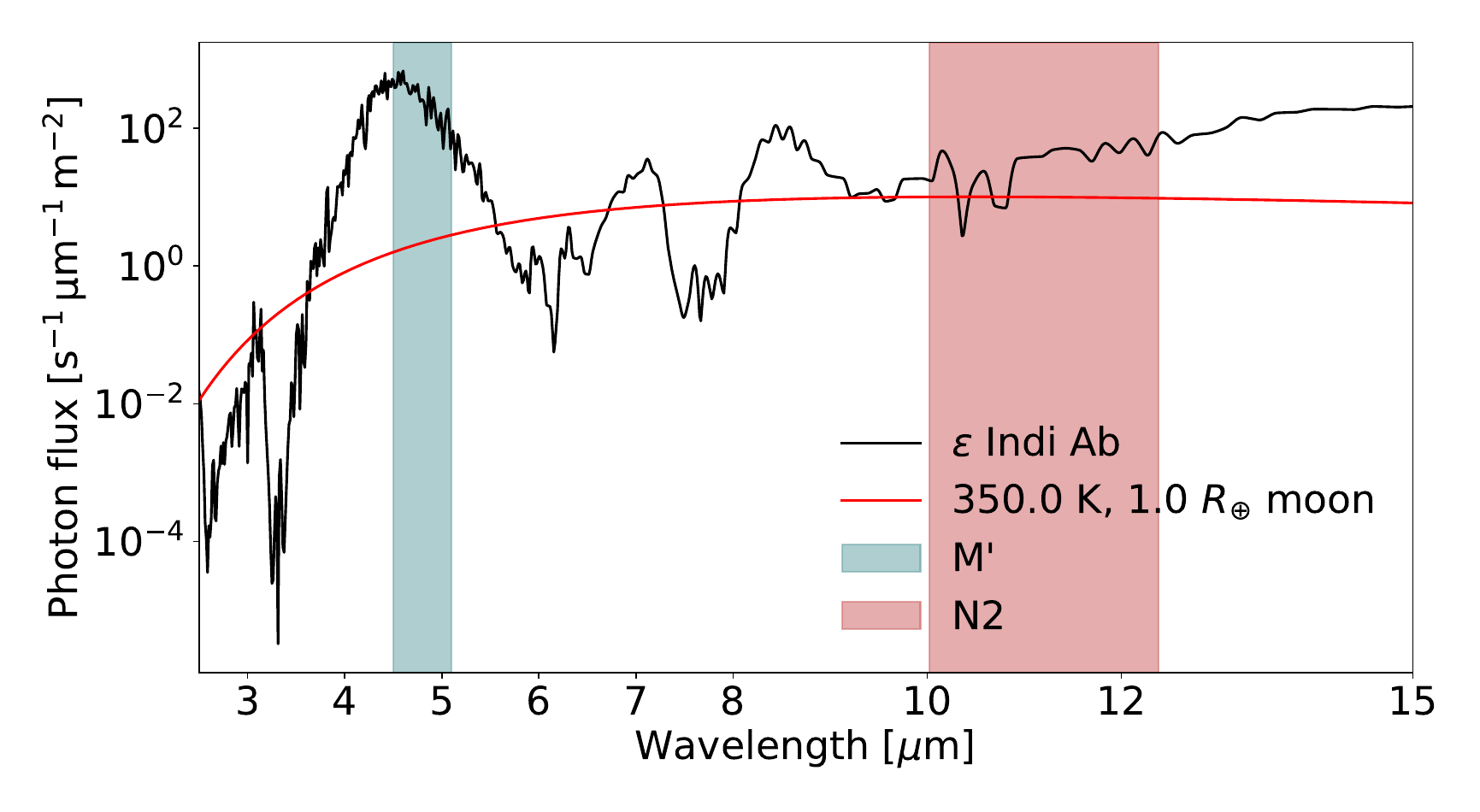}
    \caption{Spectrum for $\varepsilon$ Indi Ab, generated from \textit{ATMO2020} assuming a $3.25$ M$_{\textrm{Jup}}$, 3 Gyr planet, and equilibrium chemistry; the chosen $M'$ and $N2$ bands are marked. Also shown is the extreme scenario of a $350$ K, Earth-sized blackbody moon, for which it is clear that the planet still dominates by $\sim2$ orders of magnitude in $M'$, while the moon contributes an appreciable fraction of the flux in $N2$, such that a spectroastrometric signal may be observed between the two filters. We note that the region from 5.5-7.5 $\mu$m, though it has promising absorption regions, is inaccessible from the ground.}
    \label{fig:spectrum}
\end{figure}

To assess the limits of what types of Solar System-like moons may be detectable in nearby systems in the near-future using ELT-class telescopes, we explore a benchmark scenario. We take METIS as reference instrument, given that its capability to detect Earth-sized planets around nearby stars in thermal emission has been well-established \citep{Brandl2021METIS:Spectrograph}; it should therefore not be problematic to reach a $5\sigma$ detection (in flux) of a giant planet, so as to satisfy the assumptions required for Eq.~\ref{eq:flux_noise}. The telescope model parameters are summarised in Tab.~\ref{tab:benchmark_system}. For the moon filter $M$, we take the $N2$-filter as it is the longest-wavelength continuum filter on METIS, which will therefore be best-suited to capture low-temperature ($<300$ K) thermal emission. For the planet filter $P$, by contrast, we choose the near-infrared filter $M'$, which captures a region in which giant planets such as Jupiter, Saturn \citep{Roman2023Mid-InfraredPlanets} and higher-mass exoplanets \citep{Phillips2020AExoplanets} are luminous, while remaining stiff to the types of temperatures expected globally for tidally heated moons (see e.g. \citealt{Dobos2015ViscoelasticExomoons}). As an example, the used spectrum for $\varepsilon$ Indi Ab (which we shall justify shortly) is shown with a $350$ K, Earth-sized moon (representing an upper bound on the hottest, largest plausible moons) for comparison in Fig~\ref{fig:spectrum}: it is clear that the planet still dominates wholly in $M'$, while the moon has an appreciable contribution in $N2$. METIS requirements mandate that its optical system is diffraction-limited \citep{Brandl2021METIS:Spectrograph}, which we will therefore take as an assumption on the PSF-width. Atmospheric transmission is set to an achromatic 80\%, based on a representative scenario simulated using ESO's \textit{SkyCalc} \citep{Jones2013AnParanal, Noll2012AnRange}. We set an observation time of 6 hours, representing a length of night that can be reasonably expected to occur regularly at the ELT site \citep{Lombardi2009TheObservatory}, though we note that with a scaling of the noise with $T^{-1/2}$, our results are relatively stiff to the total observation time.

We model a reference moon-planet system based on $\varepsilon$ Indi Ab, the nearest Jupiter analogue known imageable by JWST (and thus, presumably, ELT), and its nominal parameters as determined by \citet{Feng2019DetectionData}. The parameters necessary for this analysis are given in Tab.~\ref{tab:benchmark_system}; based on this planet mass and age, a spectrum is interpolated from the chemical equilibrium atmosphere-spectra in the publically available atmosphere library \textit{ATMO2020} by \citet{Phillips2020AExoplanets}. As their simulated atmospheres only provide data for planets of this mass up to ages of $\sim 3$ Gyr, we cannot take an age in the age range of $3.7-4.3$ provided by \citet{Feng2019DetectionData}, though we note that preliminary results using \textit{ATMO2020} show that in this case the spectroastrometric signal-to-noise ratio is increasing with age of the planet (i.e. with decreasing temperature): this makes an age of 3 Gyr conservative.

In imitation of Solar System moons (with the notable exception of Titan), we assume the moon to be an airless icy or rocky body. In such a case, the observed brightness temperature in the infrared agree relatively well with the surface temperature (to within a couple $\sim10$ K), regardless of surface material (see e.g. \citealt{Hu2012TheoreticalSurfaces, Whittaker2022The3844b}); integrated brightness temperatures of Jupiter's icy moons Ganymede and Callisto agree relatively well with their expected surface temperatures \citep{Squyres1980SurfaceCallisto}. We therefore choose to model the moon as a black body, such that the only free parameters that remain to explore are (1) its size (but notably not directly its mass), (2) its surface temperature and (3) its orbital properties. Each of these we vary over ranges made plausible by a combination of observed Solar System moons and moon formation theory, so as to explore the signal-to-noise ratio of the spectroastrometric signal. In particular, we vary the size between an Io-radius and an Earth-radius, which, allowing the satellite to be either icy or rocky, covers the full range of expected moon masses of $10^{-5}-10^{-3}$ host masses for large satellites proposed in literature (e.g. \citealt{Canup2006APlanets, Cilibrasi2018SatellitesMoons, Moraes2018GrowthDisc}). We vary the semi-major axis over the range seen for the Galilean satellites.

While analysis of time-series observations of the system under the assumption of a moon on a Keplerian orbit may allow one to produce higher-significance detections (see Sec.~4.2.2 in \citet{Agol2015THEEXOMOONS} for an example), we prefer to analyse solely the spectroastrometric signal in one observation, as it requires no assumptions on the nature of the signal. A high-significance detection of a spectroastrometric signal would therefore provide incontrovertible evidence of a signal of astrophysical origin, regardless of whether it is a moon or not; further analysis and conclusions on the nature of this signal can then follow.

\begin{figure*}[t]
    \centering
    \includegraphics[width=\textwidth]{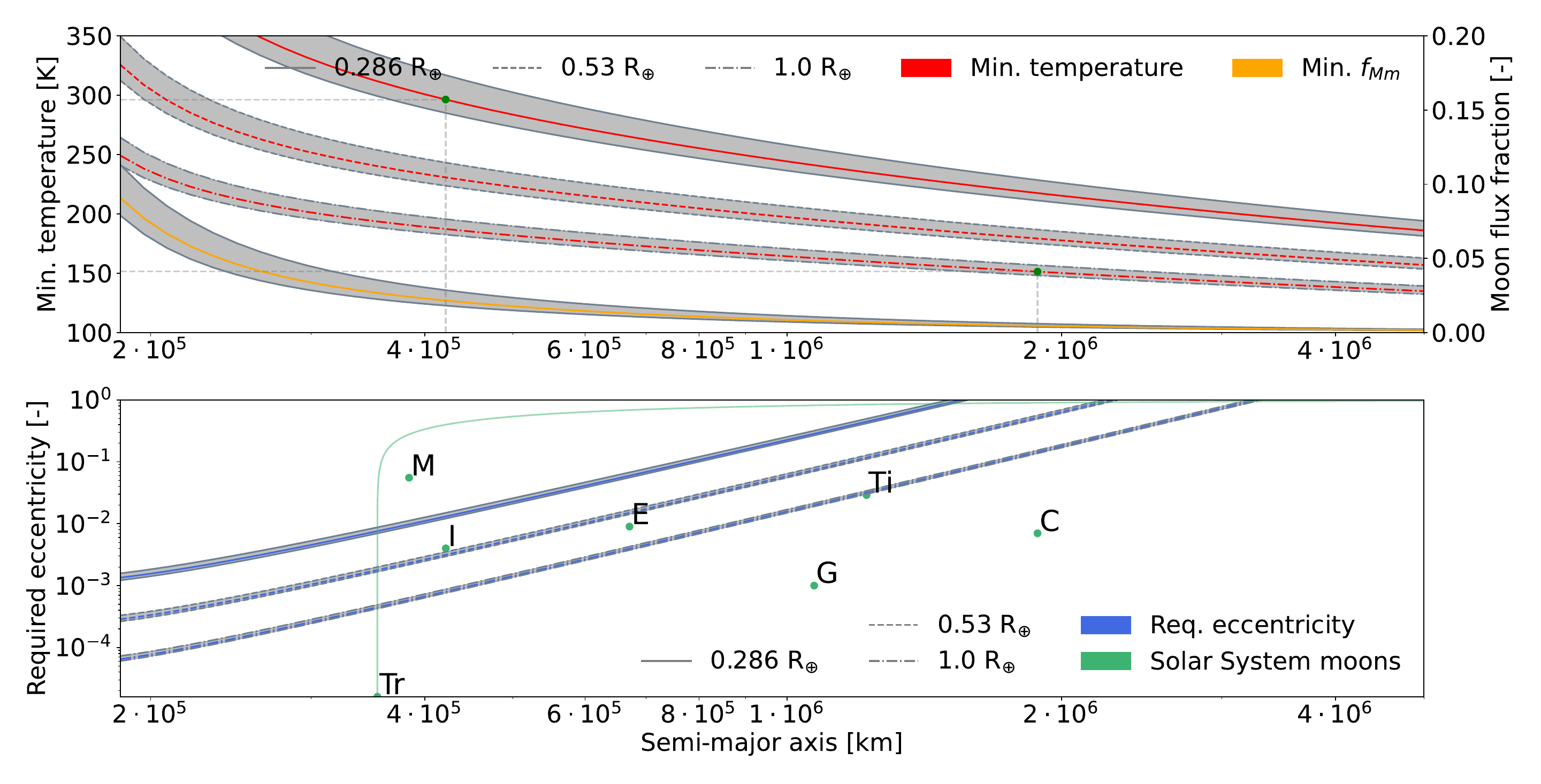}
    \caption{Minimum blackbody temperature and corresponding moon flux fraction in $M$ for a total observation time of 6 hours on ELT/METIS as a function of semi-major axis for $5\sigma$ detectability of a moon around $\varepsilon$ Indi Ab for an Io-sized, Mars-sized, and Earth-sized moon (top), with a corresponding first-order estimate of the required eccentricity (bottom). The grey region delineates the area bounded by the edge-on and face-on cases ($p=2/\pi$ and $p=1$); the red and blue lines correspond to the inclination-averaged case ($p\approx0.842$). Semi-major axes and eccentricities of the seven largest Solar System moons Ganymede, Titan, Callisto, Io, the Moon, Europa and Triton are marked by their first letter(s). The post-capture migration path of Triton assuming conservation of angular momentum (such as in \citealt{Ross1990TheTriton}) is also marked to illustrate the observational potential of captured moons. The moons used for further analysis in Secs.~\ref{subsec:results_eccentricity} and \ref{subsec:results_distance} are marked in green in the top plot. For all three moons over all explored semi-major axes, the minimum $5\sigma$-detectable temperature corresponds to a spectroastrometric signal of roughly 0.013 mas.}
    \label{fig:min_temp_a}
\end{figure*}

\section{Results} \label{sec:results}
We structure our results as follows; we examine the minimum temperature required for detectability in circular orbits as a function of semi-major axis in Sec.~\ref{subsec:results_temp}, and perform a first-order analysis of the required eccentricities for these temperatures under the assumption that the detectability is only marginally affected by low eccentricities. We validate this assumption in Sec.~\ref{subsec:results_eccentricity}, and finally perform an analysis of the attainable signal-to-noise ratios as a function of moon blackbody temperature and distance for two sample moons of given size and semi-major axis in Sec.~\ref{subsec:results_distance}.

\subsection{Minimum moon temperature for detectability}
\label{subsec:results_temp}
Fig.~\ref{fig:min_temp_a} shows the minimum required blackbody temperature of a moon for various sizes (corresponding to Io, Mars and the Earth) to be detectable at $5\sigma$ in a circular orbit around $\varepsilon$ Indi Ab, assuming a total of 6 hours of observation time. Also shown is a first-order estimate of the eccentricities required for each of the various size bodies at each semi-major axis to reach the given minimum temperature by assuming radiative equilibrium of a blackbody moon experiencing viscoelastic dissipation as given by \citet{Segatz1988TidalIo} and using the value of -Im($k_2$) given for Io by \citet{Lainey2016QuantificationData}.
For comparison, the semi-major axes and eccentricities of the seven largest Solar System moons are marked; as Triton currently has a circular orbit but is known to have migrated to its current position after capture over a timespan of $\sim 1$ Gyr \citep{Ross1990TheTriton, McKinnon2014Triton}, its constant-angular momentum post-capture evolution as described by \citet{Ross1990TheTriton} is also drawn. The value of $f_{Mm}$ corresponding to these minimum temperatures is also shown; over the explored semi-major axis values, $f_{Pm}$ for the minimum detectable temperatures is negligible, and so the required values of $f_{Mm}$ coincide. For this same reason, the spectroastrometric noise is nearly constant across all semi-major axes at $\sim0.0026$ mas, corresponding to a minimum detectable spectroastrometric signal of $\sim 0.013$ mas.

\begin{figure*}[t]
    \centering
    \includegraphics[width=1.0\textwidth]{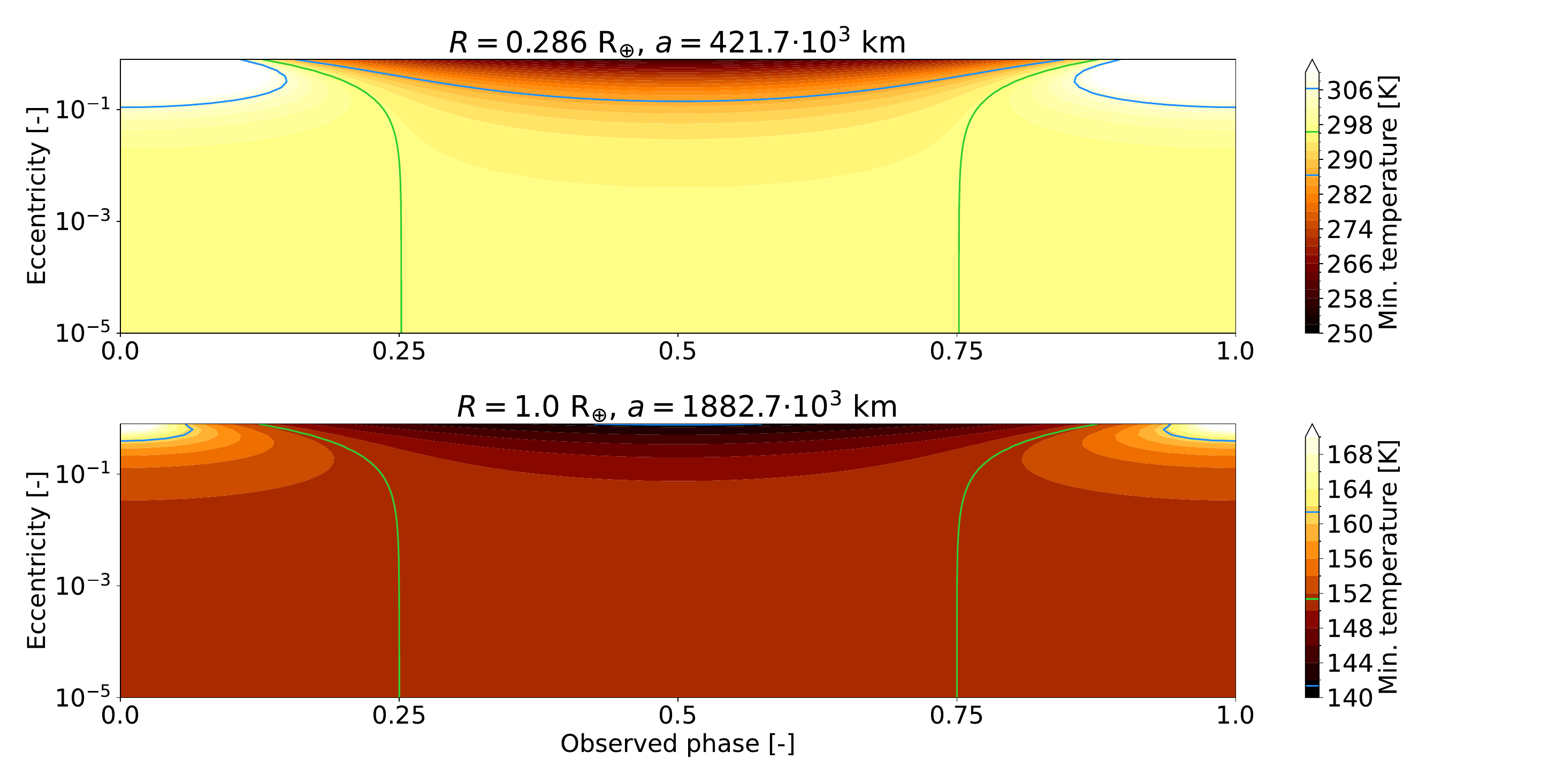}
    \caption{Minimum temperature for $5\sigma$ detectability as a function of the observed orbital phase and eccentricity for an Io-like moon with an Io-like semi-major axis (top) and for an Earth-like moon with a Callisto-like semi-major axis (bottom). The green and blue contours indicate the contour corresponding to the nominal (zero-eccentricity) required temperature (as marked on the top plot in Fig.~\ref{fig:min_temp_a}), and the 10 K deviations from this temperature, respectively; in both cases, these are also marked on the colourbar. We note that in all non-zero eccentricity cases the temperature required for observability is in fact lower than for the circular case over $>50\%$ of the orbit.}
    \label{fig:T_min_e_phase}
\end{figure*}

\subsection{Eccentricity effects}
\label{subsec:results_eccentricity}
Fig.~\ref{fig:T_min_e_phase} shows the minimum moon blackbody temperature for two sample moons (that will be explored and justified in further detail in Sec.~\ref{subsec:results_distance}) at an $\varepsilon$ Indi Ab-like system-observer distance at a variety of eccentricities and orbital phases covering all possibilities for the inclination-averaged case ($p\approx 0.842$). The green and blue lines, indicating the minimum required temperature for $e=0$ and $\pm 10$~K deviations from that contour, show that the required temperature for moons with Solar System-like eccentricities ($e\lesssim 0.1$) is in general well-approximated by the required temperature for the circular case throughout the orbit; additionally, the required temperature for detectability is in fact in all cases lower than for the circular case over $>50\%$ of the orbit.

\subsection{Applicability to distant systems}\label{subsec:results_distance}
The results presented in Secs.~\ref{subsec:results_temp} and \ref{subsec:results_eccentricity} suggest that there is a set of nearby systems in which planet-like moons of directly imageable giant planets can be detected. This motivates us to explore the extent of this viable detection space for two scenarios, representing the end-points of what \citet{Lazzoni2022DetectabilityDwarfs} deem the class of planet-like satellites (namely those formed by core accretion in the CPD, analogous to most Solar System satellites): (1) an Io-sized moon at Io-like separation from its host and (2) an Earth-sized moon at a Callisto-like separation from its host. The first requires the least number of assumptions, as it is a type of moon observed in the Solar System, and tidal heating-mechanisms through which it might reach temperatures to be luminous in the IR over observable timescales are well-established (e.g. \citealt{Dobos2015ViscoelasticExomoons, Rovira-Navarro2021}). The second would be on the larger end of what one expects to see around a host like $\varepsilon$ Indi Ab according to moon formation studies like \citet{Canup2006APlanets}, though more recently \citet{Cilibrasi2018SatellitesMoons} produced results that seem to suggest such masses may be attainable. Additionally, where fixed-Q tidal theory would predict that such a far-out moon should be unlikely to experience significant tidal interactions, the recently proposed paradigm of resonance locking would allow such far-out moons to experience tidal interactions still \citep{Fuller2016ResonanceSystems, Lainey2020ResonanceTitan}. These two cases should therefore in principle bracket the full range of plausible planet-like moons that might be expected to be luminous in the IR; the signal-to-noise ratio as a function of blackbody temperature and system-observer distance for these two cases are illustrated in Fig.~\ref{fig:SNR_T_d}.

\section{Discussion} \label{sec:discussion}
\begin{figure*}
    \centering
    \includegraphics[width=1.0\textwidth]{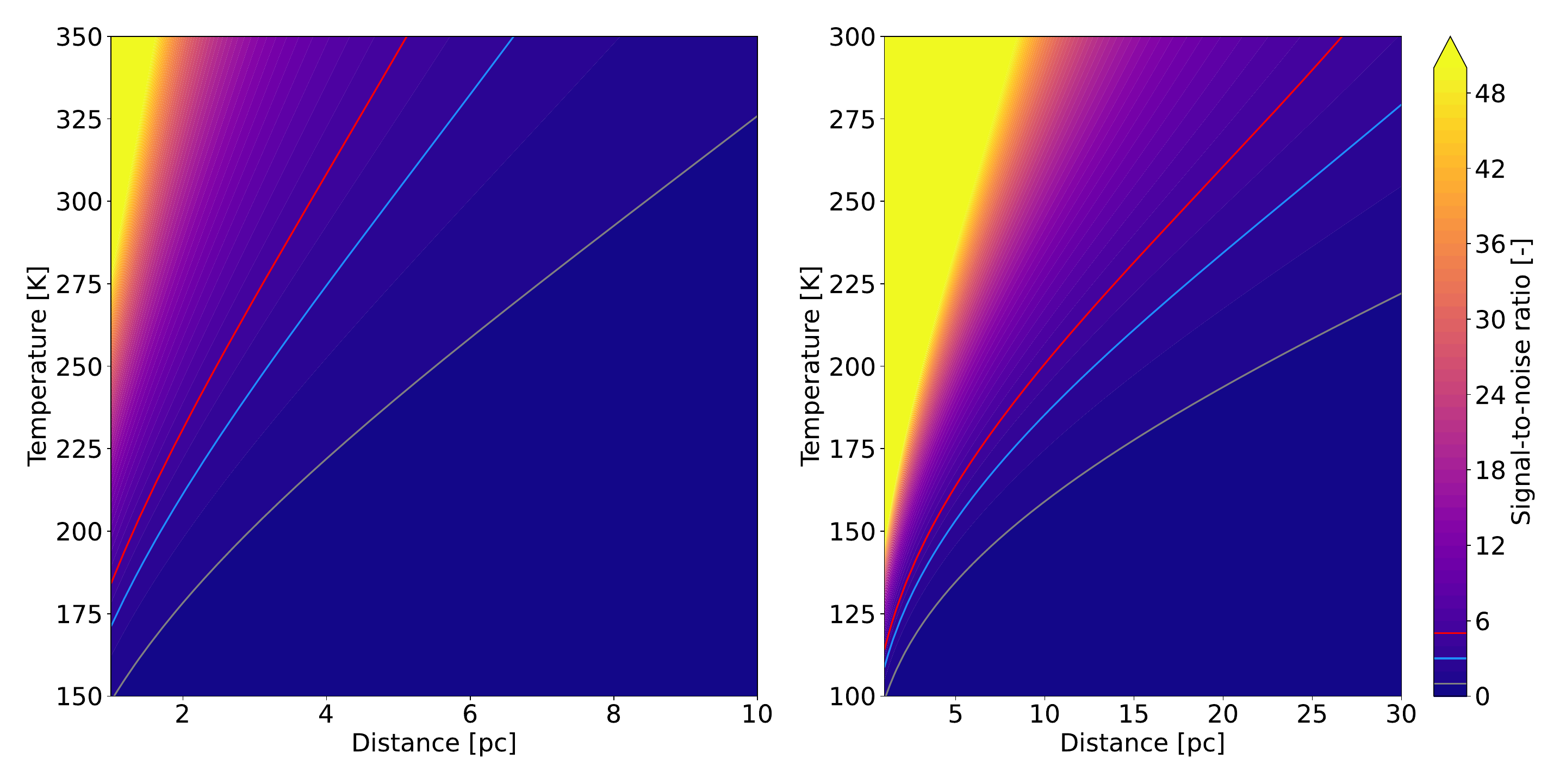}
    \caption{Signal-to-noise ratio for an Io-sized moon on an Io-like orbit (left) and an Earth-sized moon on a Callisto-like orbit (right) around an $\varepsilon$ Indi Ab-like planet as a function of blackbody temperature and system-observer distance (we note that the axis scale and datum differ between the two moons). The grey, blue, and red lines mark the $1\sigma$, $3\sigma$, and $5\sigma$ boundaries, respectively.}
    \label{fig:SNR_T_d}
\end{figure*}

Figs.~\ref{fig:min_temp_a} and \ref{fig:SNR_T_d} indicate that spectroastrometry may be able to provide detections of large icy moons (with surface temperatures as low as $T_{s}\lesssim 150$ K) or even smaller, hot rocky moons (with surface temperatures 
as low as $T_{s}\lesssim 300$ K) in nearby systems; such temperatures are plausible for tidally heated moons on orbits comparable to modern-day Solar System moons (e.g. \citealt{Dobos2015ViscoelasticExomoons, Forgan2016ExomoonHeating, Rovira-Navarro2021}) and easily exceeded by conditions as have been presumed to exist in the early Solar System (e.g. \citealt{Ross1990TheTriton}), potentially over Gyr timescales \citep{McKinnon1990TritonsHistory, Lunine1992ATriton}. The possibility of observing large icy moons at low temperatures is interesting, as the large icy moons Ganymede and Callisto in our Solar System have (and are stable at) surface (and brightness) temperatures of order $\sim150$ K \citep{Squyres1980SurfaceCallisto} (though a large part of this is contributed by solar flux), and the largest and most massive moons are thought to form at or beyond the ice line of their CPD \citep{Heller2015WATERPLANETS}, therefore being icy. The results for smaller hot moons are also of interest, as \citet{Rovira-Navarro2021} have shown that plausible scenarios may arise in which such temperatures are maintained for timescales $\gtrsim 1$ Gyr.

We will discuss these results in several contexts: in Sec.~\ref{subsec:discussion_comparison_previous}, we compare our framework against that in previous literature, and we discuss several caveats to our new formulation in Sec.~\ref{subsec:limitations}. One may also wonder if perhaps current-generation infrared telescopes may be able to detect objects in this manner, too: we discuss what this would take in Sec.~\ref{subsec:current_gen}, and follow this up with a discussion on the applicability of spectroastrometry to moon-planet systems other than the ones we have studied in this paper in Sec.~\ref{subsec:detectability_of_other_systems}. Finally, we make a comparison of spectroastrometry against other methods in Sec.~\ref{subsec:comparison_methods} and discuss what combinations with other methods may possibly allow unambiguous detection, characterisation and confirmation of moon candidates in Sec.~\ref{subsec:discussion_synergies}.

\subsection{Comparison to previous work}
\label{subsec:discussion_comparison_previous}
While we have expanded the framework provided by \citet{Agol2015THEEXOMOONS} to work for closed Keplerian orbits in general (Sec.~\ref{subsec:kepler_orbit_signal}) and to include additional noise bounds (Sec.~\ref{subsec:noise_sources}), a great deal of conclusions in their work still hold. In particular, the scaling of the signal-to-noise ratio as $S/N\propto (\epsilon T)^{1/2}d^{-2}$ remains generally applicable, and under the assumptions that the detector pixel size is designed to sample the PSF well (i.e. $\alpha \propto \sigma_{PSF}$), that the pointing stability noise is negligible and that the telescope is diffraction-limited we also recover the telescope-diameter proportionality $S/N \propto D^2$. Additionally, while \citet{Agol2015THEEXOMOONS} do not appear to explicitly have taken note of this, (1) in their formulation the noise is a function of the spectral properties of the moon, but not of its orbit (excepting transits), (2) the spectroastrometric signal-to-noise ratio of a single observation can be straightforwardly calculated for a given observation without requiring any a priori knowledge on the nature of the observed system and (3) there exists an optimal time allocation that minimises the noise of an observation that can be well-approximated by quantities that can be known or estimated a priori (Sec.~\ref{subsec:obs_time}). Each of these conclusions still hold in our expanded formulation.

There are several additions we have gained over this previous work, however. We have shown that the viability of spectroastrometry is relatively insensitive to the orbital parameters of an exomoon with the exception of its semi-major axis and particularly high eccentricities (well exceeding those found in the Solar System). We found an expression for the best-case and worst-case inclinations as well as the expected spectroastrometric signal for a flat prior on moon inclinations. Additionally, one can show from Eq.~\ref{eq:xi} that in the limit of large $T/P$, $\hat{\boldsymbol{\xi}}\to~-~[3e/2,~0]^T$, such that circular or small-eccentricity orbits become invisible. Further analysis of Eq.~\ref{eq:xi} (e.g. by analysing the pre-factor $P/(\pi T) \sin\left((E_1-E_0)/2\right)$ in Eq.~\ref{eq:xi_app}) shows that this already occurs for $T/P\gtrsim 1$. Hence, we should not expect to be able to observe moons with periods on the order of or shorter than our observation time in the relevant filter, both because their semi-major axis and the resulting orbital motion is small, which was already apparent from the work by \citet{Agol2015THEEXOMOONS}, but also because their signal is washed out by this time-averaging effect for $T/P\gtrsim1$. Conversely, highly eccentric moons on short-period orbits may instead become visible precisely \textit{because} of this time-averaging effect. While for ground-based telescopes observation times are unlikely to exceed orbital periods, this is a plausible scenario for space-based observations, which would require far longer observation time to reach a performance similar to large-diameter ground telescopes.

Additionally, we have shown that while background and instrument noise in centroid measurements for a significant ($>5\sigma$) detection of the planet is likely to be of similar magnitude to the photon noise, the noise due to a pixelated detector (for a well-sampled PSF) and pointing stability of the telescope in question are unlikely to provide major contributions to the centroid noise, and therefore to the spectroastrometric noise.

\subsection{Limitations on this formulation}
\label{subsec:limitations}
A major limitation of the method as developed in this manner is the lack of an expression for speckle noise, which ostensibly does not satisfy the assumption of independence between pixels required for Eq.~\ref{eq:flux_noise}. Until this effect can be quantified, these results should be taken only as representative for moons of far-out or free-floating planets, where speckle noise is negligible, though fortunately the majority of large exomoons are expected to occur around wide-orbit planets \citep{Dobos2021SurvivalExoplanets, Heller2015ConditionsStars}, beyond the snowline \citep{Inderbitzi2020FormationInstability}. If an expression or estimate for speckle noise in spectroastrometry can be found, we might be able to push spectroastrometry to find moons of close-in planets in reflected light, too, like the case of an Earth-Moon analogue studied by \citet{Agol2015THEEXOMOONS}, though this would require a different (augmented) model of the moons' spectral energy distribution incorporating reflected light.

A second limitation for close-in planets is the lack of inclusion of the planet-moon barycentre movement over the observations in our formulation. While the movement thereof between observations can be accounted for if the orbital properties of the planet about its host are known, the movement throughout the observation will likely have an effect that is worth quantifying in future studies: using the orbital parameters given by \citet{Feng2023RevisedJupiters}, the worst-case angular velocity (i.e. for a face-on orbit at pericentre) for $\varepsilon$ Indi Ab is on the order of 0.1 mas/hr, which is of the same order as spectroastrometric effects. The authors deem it likely that this can be accounted for in a similar fashion to the time-averaging method employed in App.~\ref{sec:angular_separation_kepler}, either numerically or possibly analytically, by inserting the barycentre motion into the integral as an offset. For wide-separation planets or edge-on planets observed at their most distant apparent separation, this effect is negligible, however. As our neglect of speckle noise means that wide-separation planets should be preferred to start with, for this effect to become strong will most likely require that speckle noise be dealt with, first.

Another matter that needs mentioning is the fact that there are objects that cannot immediately be distinguished from moons; background objects may, per chance, induce a spectroastrometric signal. Follow-up observations will, however, reject such candidates without issue, given the rapid motion expected of moons about their host that such background objects will not display. Additionally, thermal emission from asymmetric ring systems around planets will cause a signal that may mimic thermal emission from a hot moon; one could foresee such asymmetric rings arising in a miniature version of the scenario invoked by \citet{Andras2020NewFomalhaut} for the debris cloud formerly identified with Formalhaut b, for example, though notably Formalhaut b was not detectable in the infrared. As such asymmetries on a Keplerian orbit should be expected to dissipate over the order of several orbits, a signal due to such occurrences should not be expected to remain stable over observational timescales; additionally, the presence of rings (asymmetric or not) may be excluded altogether by the method proposed by \citet{Lazzoni2020TheB}, for example. 

Spots or asymmetric patterns on the planet may produce an astrometric signal, too: we note that the Great Red Spot on Jupiter is dim in the infrared \citep{Ge2019RotationalExoplanets}, and so we should by analogy expect any such effects to stem from larger-scale variability. For Jupiter, such rotational variability is on the order of 1\% in flux around the 10.5 $\mu$m region, whereas in the 5 $\mu$m region this is roughly 20\% \citep{Ge2019RotationalExoplanets}. For the moon filter we can thus safely disregard this variability: the planet filter does warrant a small calculation. In the worst-case scenario of a pole-on planet with all this variability concentrated on a single spot at the equator, and with a Jupiter-like radius and rotation period of 10 hours, the resultant astrometric signal will be on the order of $10^{-2}$ mas. As the 20\% variability on Jupiter is the result of large-scale variability, not spot-like localised variability \citep{Ge2019RotationalExoplanets}, we expect that this effect will be significantly smaller in practice. Additionally, higher-mass planets are expected to have a lower rotational period \citep{Snellen2014FastB}, which will dampen the signal strongly if the rotational period becomes of the same order as the observation time in a filter (see Eq.~\ref{eq:circular_average_orbit}).  Nonetheless, these variability effects may thus, in the worst case, be present at an order of magnitude that will contaminate the most sensitive of our results: follow-up observations in different planet-bands (perhaps with lower variability) will then provide a conclusive result on the presence or absence of a moon.

It is thus possible to exclude background objects, (asymmetric) ring systems, or planet variability either immediately or after follow-up observations. Consequently, we expect that moons are the only objects or phenomena that can be responsible for long-term spectroastrometric signals, though any putative signal will need follow-up observations for confirmation.

\subsection{Spectroastrometric capabilities of current-generation telescopes}
\label{subsec:current_gen}
As these results show that observing tidally heated exomoons on future IR telescopes is plausible, one might wonder whether current-generation IR telescopes have the capabilities required to observe tidally heated moons in nearby systems. The prime candidate here, of course, is JWST; unfortunately, the scaling of the signal-to-noise ratio with $D^{2}$ suggests that the observation time required would be on the order of days. JWST would thus only be sensitive to hot moons around nearby directly imageable planets that are less luminous than $\varepsilon$ Indi Ab, none of which are currently known.

Additionally, while the fine-guidance pointing stability of JWST is of an order similar to that of ELT, target reacquisition (which would be required when switching between filters) only has accuracies on the order of mas \citep{Hartig2022JWSTCommissioning}, which introduces systematic error. This may be remedied if one uses astrometric observations (i.e. using a single filter) of the moon throughout its orbit instead, but in that case it still remains to be shown that the measured effect is due to motion of a moon rather than due to a third object disturbing the Keplerian motion of the planet. In fortituous circumstances, one might also be able to use sufficiently bright background objects (that can be astrometrically positioned to the required accuracy) to reference the observations in the two filters against each other, but this requires that such objects be available and fixed. We leave it to future work to determine the viability of each of these methods to perform spectroastrometric measurements using JWST.

\subsection{Detectability of other moon-planet systems}
\label{subsec:detectability_of_other_systems}
In this analysis, we have considered Solar System-like moons around a planet analogous to $\varepsilon$ Indi Ab, as it is the nearest known Jupiter analogue that is directly imageable. However, such a system need not be optimal in terms of detectability of any moons; moreover, it is likely that with JWST operational many more directly imageable planets will be discovered in nearby system that had previously eluded detection. It is thus productive to consider what other planets might be found, or perhaps what types of moons might be detectable around them.

One such category is lower-mass planets; exploratory results using the code in this paper suggest that even lower-temperature moons will be observable around lower-mass planets, as the relative flux of the moon then increases appreciable in the moon filter. If located in a suitable nearby system (such that stellar contamination is not a problem), this may thus allow the detection of true Solar System-analogues: that is to say, Solar System-analogue planets accompanied by Solar System-size and Solar System-temperature moons. Down to Saturn-masses this is hardly problematic, but for lower-mass planets (transitioning into the ice giant-regime) the spectral feature at $5$ $\mu$m starts to become muted \citep{Roman2023Mid-InfraredPlanets, Linder2019EvolutionaryDetectability}, such that the planet will no longer outshine its moon in that band; another planet filter than M' (which was selected in a fairly ad hoc fashion) is required for ice giants. Detection of moons around such planets would thus require a more robust filter selection procedure than we have followed in our analysis of $\varepsilon$ Indi Ab. Hence, we propose that in anticipation of the detection of directly imageable low-mass giant planets in nearby systems by JWST further research is done on filter performance over a wide variety of planet masses and ages.

A second category of systems does not concern the type of planet, but rather the type of moon; while we have for the purposes of our analysis explicitly remained in the familiar realm of Solar System-like moons, current exomoon candidates are ostensibly unlike the moons seen in our Solar System. The candidates put forward by \citet{Teachey2018HEK.I}, \citet{Teachey2018EvidenceKepler-1625b}, \citet{Lazzoni2020TheB} and \citet{Kipping2022AnB-i} are perhaps rather thought of as binary objects; this poses unique challenges for spectroastrometry in that the spectra of two gas giant planets in a mutual orbit are likely to posses similar features qualitatively over a broad range of wavelengths. As both objects are, without requiring tidal heating, likely to be relatively bright, however, their spectroastrometric signal may still be detectable, and if so, perhaps at system distances further away than Solar System-like moons. Given the importance of such objects in informing moon formation theory (e.g. \citealt{Hamers2018I, Hansen2019FormationCapture, Moraes2020ExploringI}), it is worth exploring whether any filter combination might be able to detect or rule out the presence of any such moons; perhaps, even, whether a filter combination may be able to detect both Solar System-like moons and binary-like objects.

\subsection{Comparing against other moon detection methods}
\label{subsec:comparison_methods}
Over other methods, spectroastrometry provides an additional boon in the fact that, if the barycentre motion discussed in Sec.~\ref{subsec:limitations} is accounted for, the astrometry produced is of a sufficient level to potentially detect the orbital motion of nearby exoplanets over a single observation; even in the absence of a moon detection, the data is therefore still useful in novel ways. Another major advantage that spectroastrometry has over the current suite of standard exomoon detection efforts is the lack of degeneracies (if observed over multiple epochs), such as those induced by unseen second planets in the case of the methods proposed for transiting planets \citep{Fox2021ExomoonExomoons, Kipping2020IMPOSSIBLEEXOMOON}; as one measures explicitly the movement of the moon about its planet, these third-object effects are unambiguously removed.  Finally, spectroastrometry provides the possibility of repeatable observations of nearby moons as small and nearly as cold as those observed in the Solar System, far below temperatures required for other direct imaging methods for tidally heated moons (e.g. \citealt{Kleisioti2023TidallyCharacterization}), while being relatively insensitive to the inclination or orientation of the moon or its planet.

\subsection{Synergies with other methods}
\label{subsec:discussion_synergies}
Further modelling of the moon or its orbit can further enhance the significance of results and produce predictions or constraints on quantities such as the semi-major axis, period, size, host planet mass and flux distribution between satellite and host \citep{Agol2015THEEXOMOONS} and consequently even interior properties of the satellite (e.g. \citealt{Kleisioti2023TidallyCharacterization}) that can be (independently) verified with other methods, such as radial velocity or astrometric measurements of the planet or such as TTVs/TDVs, auxiliary stellar transits or moon-transits of the planet, if the system is fortituously oriented. Especially noteworthy is the fact that at satellite inclinations where spectroastrometry fares poorest (near edge-on), transits of their host and radial velocity measurements become viable, which makes spectroastrometry a useful complement to those methods. Finally, mutual events \citep{Cabrera2007DetectingEvents} observed in the infrared might provide information on the composition of the planet and moon as well as provide independent constraints on the moon orbit \citep{Schneider2015AExo-moons}.

Another detail of particular interest is the fact that spectroastrometry provides an independent constraint on moon and planet fluxes as well as the moon orbit that may be combined with those derivable from moon-planet models such as discussed in \citet{Kleisioti2023TidallyCharacterization}, which can be produced from the same direct imaging data that spectroastrometry can be performed on. In combination with the results derivable from other methods, it is therefore possible to come to a near-full characterisation of the moon in question in terms of its orbit, composition, size and surface conditions.

We would therefore like to emphasise the possible capabilities of simultaneous use of spectroastrometry over multiple epochs, photometric modelling, astrometry and transit observations of moon-transits of the planet, as all of these should in principle be possible with the same series of direct-imaging observations. Given the set of mutually independent estimates for similar parameters available between the set of these, self-consistency of any candidate can be straightforwardly checked. We therefore strongly recommend that the viability of this combination is evaluated in future studies.

\section{Conclusion} \label{sec:conclusion}
Spectroastrometry has previously been shown to be a promising method for detection of Earth-Moon-like systems or Earth-like moons orbiting Jovians by \citet{Agol2015THEEXOMOONS}. We have shown that there is a further class of satellites, tidally heated moons analogous to those in the Solar System, that are observable in nearby systems using this method with next-generation ground telescopes. Illustrated by two example systems motivated by moons observed in our current-day Solar System, we see that for nearby systems even large icy bodies or hot bodies comparable in size to those seen in our Solar System may be observable.

In showing this, we have derived an expression for the spectroastrometric signal that covers all closed Keplerian orbits, along with the orbital motion of the moon throughout the observation, which shows that the efficacy of spectroastrometry is only weakly dependent on the orbital properties of an exomoon, except for moons with periods equal to or lower than the observation time, which are unlikely to be detected. Moreover, we have derived additional conservative noise estimates for noise due to (1) background and instrument noise, (2) a pixelated detector and (3) pointing inaccuracies. This now allows evaluation of spectroastrometry as a method on telescopes other than the ideal photon-noise limited space telescopes assumed in previous research and without requiring a priori any assumptions on the orbit or nature of any potential moon.

\emph{Acknowledgements.} We would like to thank Amy Louca and Christiaan van Buchem for the productive discussions on the prospects of detecting tidally heated moons and Neptune-like moons that led to the research presented in this paper, and Matthew Kenworthy for motivating us to work towards publishing the result. We must also acknowledge the insightful comments provided by the anonymous referee, which have greatly strengthened the presented discussion of spectroastrometry.

\bibliographystyle{aa}
\bibliography{references_from_mendeley}

\begin{appendix}
\section{Derivation of the general time-averaged angular separation between moon and planet for closed Keplerian orbits} \label{sec:angular_separation_kepler}
\noindent Let us suppose that we take an observation over a time period $[t_0, t_0 + T]$ of a moon-planet system in mutual (closed) orbit around one another with Keplerian elements $(\alpha, i, e, \Omega, \omega, M_0)$ at a system-observer distance $d$. Here we use the astronomical conventions: the inclination is measured between the orbital angular momentum vector and the line-of-sight (such that the reference plane is oriented perpendicular to the line-of-sight) and the longitude of the ascending node is measured from celestial north eastward (i.e. anti-clockwise). To predict the measured spectroastrometric signal (Eq.~\ref{eq:spectroastrometric_signal}) we then need the time-averaged on-sky position of the moon with respect to its host, $\left<\mathbf{c}_{mp}(t)\right>$: as the on-sky projection of this time-averaged position are precisely the $x$- and $y$-components in the reference coordinate system (i.e. the coordinates in the plane perpendicular to the line-of-sight along celestial north and east, respectively), we have:
\begin{align}
    \left<\mathbf{c}_{mp}(t)\right> &= \frac{1}{T}\int_{t_0}^{t_0+T} \mathbf{c}_{mp}(t) \dd{t} = \frac{1}{Td}\int_{t_0}^{t_0+T}r_{mp}(t)\dd{t} \nonumber \\ &=\frac{1}{d}\mathrm{R}_3(\Omega) \mathrm{R}_1(i) \mathrm{R}_3(\omega)\frac{1}{T}\int_{t_0}^{t_0+T} 
    \begin{pmatrix}
    \xi \\
    \eta
    \end{pmatrix} \dd{t}
\end{align}
where $\xi$ and $\eta$ denote the coordinates in the orbital plane with $\xi$ measured from barycentre to pericentre and $\eta$ oriented along the latus rectum so as to have a right-handed frame; moreover, $\mathrm{R}_i(\alpha)$ denotes the rotation matrix for the rotation by an angle $\alpha$ about the $i$-th coordinate axis; these rotations can be pulled out of the integral by linearity, meaning that in fact we need only calculate the average position of the moon within its orbital plane and transform this by its orbital elements to the on-sky plane in the usual way. For brevity, let us denote the integral at the end of the expression (which is in fact the position of the moon in the orbital plane averaged over time) $a\hat{\boldsymbol{\xi}}$ (including $a$ will make $\hat{\boldsymbol{\xi}}$ dependent on $a$ only through the orbital period, as we will later see), such that:
\begin{equation}
    \left<\mathbf{c}_{mp}(t)\right> = \frac{a}{d}\mathrm{R}_3(\Omega) \mathrm{R}_1(i) \mathrm{R}_3(\omega)\hat{\boldsymbol{\xi}}.
\end{equation}
From basic orbital mechanics, we observe that:
\begin{align}
    \hat{\boldsymbol{\xi}} &= \frac{1}{aT}\int_{t_0}^{t_0+T} r \begin{pmatrix} \cos{\theta} \\ \sin{\theta} \end{pmatrix} \dd{t} \nonumber \\ &= \frac{P}{T}\frac{(1-e^2)^{5/2}}{2\pi} \int_{t=t_0}^{t=t_0+T} \begin{pmatrix} \cos{\theta} \\ \sin{\theta} \end{pmatrix} \frac{\dd{\theta}}{\left(1+e\cos{\theta}\right)^3}
\end{align}
where we have used that $\dot{\theta} = \sqrt{a(1-e^2)\mu} r^{-2}$, introducing the gravitational parameter $\mu$; moreover, we have used that $r = a(1-e^2)(1+e\cos{\theta})^{-1}$ and that for the period $P$ of an orbit we have $P=2\pi\sqrt{a^3\mu^{-1}}$ (for each of these, one can consult any orbital mechanics textbook e.g. \citet{Curtis2013OrbitalStudents}). We obtain:
\begin{equation}
\label{eq:weierstrass_form}
    \hat{\boldsymbol{\xi}} = \frac{P}{T}\frac{(1-e^2)^{5/2}}{2\pi} \int_{t=t_0}^{t=t_0+T} \begin{pmatrix} \cos{\theta} \\ \sin{\theta} \end{pmatrix} \frac{\dd{\theta}}{\left(1+e\cos{\theta}\right)^3}.
\end{equation}
It is not immediately clear how to go about solving the integral in Eq.~\ref{eq:weierstrass_form}, but in general one can integrate rational integrands of trigonometric functions through the Weierstrass substitution $x=\tan{(\theta/2)}$ (see e.g. \cite{Stewart2015Calculus:Version}, p. 502). We follow a similar route, but also note that for elliptical and circular orbits we might as well immediately make use of the relation between the eccentric anomaly $E$ and the true anomaly $\theta$,
\begin{equation}
    \tan{\frac{E}{2}} = \sqrt{\frac{1-e}{1+e}}\tan{\frac{\theta}{2}}
\end{equation}
(again, see e.g. \citet{Curtis2013OrbitalStudents}. In this form we effectively perform the Weierstrass substitution $x=\tan{(\theta/2)}$ followed by the substitution $E = 2\arctan{\left(\sqrt{(1-e )/(1+e)}x\right)}$) which after some algebra reduces expression~\ref{eq:weierstrass_form} to a set of standard integrals which can be evaluated to yield:
\begin{align}
    \hat{\boldsymbol{\xi}} &= \frac{P}{2\pi T}
    \begin{pmatrix}
    (1+e^2)\sin{E} - \frac{e}{4}\sin{2E} - \frac{3e}{2}E \\
    (1-e^2)^{1/2} \left(-\cos{E} + \frac{e}{4}\cos{2E}\right)
    \end{pmatrix}\bigg\rvert_{E=E_0}^{E=E_1} \nonumber \\
    &= \frac{P}{2\pi T}
    \begin{pmatrix}
    (1-\frac{e^2}{2})\sin{E} - \frac{e}{4}\sin{2E} \\
    (1-e^2)^{1/2} \left(\frac{e}{4}\cos{2E} - \cos{E}\right)
    \end{pmatrix}\bigg\rvert_{E_0}^{E_1} -  
    \begin{pmatrix}
    \frac{3e}{2} \\
    0
    \end{pmatrix}\nonumber \\
    &= \label{eq:xi_app}\frac{P}{\pi T} \sin\left(\frac{E_1-E_0}{2}\right) \left[
    \begin{pmatrix}
        (1-\frac{e^2}{2})\cos\left(\frac{E_1+E_0}{2}\right) \\
        (1-e^2)^{1/2}\sin\left(\frac{E_1+E_0}{2}\right)
    \end{pmatrix}
     \right. \nonumber \\
     &- \left.\frac{e}{2}\cos\left(\frac{E_1-E_0}{2}\right)
     \begin{pmatrix}
     \cos\left(E_1+E_0\right) \\
     (1-e^2)^{1/2}\sin\left(E_1+E_0\right)
     \end{pmatrix}\right] \nonumber \\ &-\begin{pmatrix}
        \frac{3e}{2} \\
        0
    \end{pmatrix}
\end{align}
where we define $E_1=E(t_0+T)$ for notational convenience, and note that in fact $E_0=E_0(\tfrac{t_0}{P}, e)$ and $E_1=E_1(\tfrac{t_0+T}{P})$, such that all semi-major axis-dependency that was still contained in $P$ can be parametrised in terms of period phase. By using the eccentric anomaly-true anomaly relation rather than solely the Weierstrass substitution we have our result solely in terms of quantities from orbital mechanics that are either known or that can be obtained from Kepler's equation for which efficient, well-known solution computation methods exist (e.g. \citealt{Mikkola1987AEquation, Markley1995KeplerSolver, Fukushima1997AEVALUATIONS, Zechmeister2018CORDIC-likeEquation}).

\section{Properties of the time-averaged angular separation}
\label{sec:app_properties}
We remark that up until $e\approx0.43$, $\sqrt{1-e^2}\approx1$ to within $10\%$, and that up until $e\approx0.77$ we have $\sqrt{1-e^2}\approx 1-e^2/2$ to within $10\%$. For the vast majority of plausible eccentricities, therefore, $\hat{\boldsymbol{\xi}}$ is well-approximated (i.e. to within $10\%$) as:
\begin{align}
    \hat{\boldsymbol{\xi}} &\approx 
    \frac{P\sqrt{1-e^2}}{\pi T} \sin\left(\frac{E_1-E_0}{2}\right)
    \begin{pmatrix}
        \cos\left(\frac{E_1+E_0}{2}\right) \\
        \sin\left(\frac{E_1+E_0}{2}\right)
    \end{pmatrix}
     \nonumber \\
     &- \label{eq:app_xi_approx} \frac{P}{\pi T}\frac{e}{2}\sin\left(E_1-E_0\right)
     \begin{pmatrix}
     \cos\left(E_1+E_0\right) \\
     \sin\left(E_1+E_0\right)
     \end{pmatrix} -\begin{pmatrix}
        \frac{3e}{2} \\
        0
    \end{pmatrix}.
\end{align}
As $E_1 - E_0 \sim 2\pi T/P$ (from Kepler's equation), we see that for eccentricities $e\lesssim 0.4$ the dynamic part (comprising the first two terms) of $\hat{\boldsymbol{\xi}}$ is dominated by a term of magnitude $\sim \sqrt{1-e^2}\textrm{sinc}{(\pi T/P)}$ pointing roughly in a direction $\Tilde{E}=(E_1 + E_0)/2$, with a secondary (correction) term of order $\sim e/4 \textrm{sinc}{(2\pi T/P)}$ in the direction $2\Tilde{E}$, where we adopt the convention that $\textrm{sinc}{x} = x^{-1}\sin{x}$. We thus see that this second term is only important when $T\gtrsim P$ and $e$ is close to $1$, but also that its magnitude is never greater than the last, static component, which starts to play an important role for eccentricities of order $e\gtrsim 0.1$. A useful approximation for first-order estimate purposes (though too inaccurate for computational purposes) for $\hat{\boldsymbol{\xi}}$ is therefore given by:
\begin{equation}
    \hat{\boldsymbol{\xi}} \sim \sqrt{1-e^2}\textrm{sinc}\left(\frac{\pi T}{P}\right)
    \begin{pmatrix}
        \cos{\Tilde{E}} \\
        \sin{\Tilde{E}}
    \end{pmatrix} - \begin{pmatrix}
        \frac{3e}{2} \\
        0
    \end{pmatrix}.
\end{equation}
In the circular limit ($e=0$) this becomes exact, giving the particularly simple form:
\begin{equation}
\label{eq:circular_average_orbit}
    \hat{\boldsymbol{\xi}}(t_0, \tfrac{T}{P}) = \textrm{sinc}\left(\pi\tfrac{T}{P}\right)
    \begin{pmatrix}
        \cos{\left(\theta_0 + \pi\frac{T}{P}\right)} \\ \sin{\left(\theta_0 + \pi\frac{T}{P}\right)}
    \end{pmatrix}
\end{equation}
where we have defined $\theta_0=\theta(t_0)$.

An interesting conclusion can be drawn from these expressions. As we observe the moon at an arbitrary time (i.e. we can consider the mean anomaly at which we observe the moon to be a uniformly distributed random variable) we can compute that the expected value of the component of $\left<\mathbf{c}_{mp}(t)\right>$ along the major axis of the orbit has magnitude $\frac{3e}{2}\frac{a}{d}$ (that along the latus rectum, as one should expect given symmetry, is zero), regardless of the observation time. In fact, in the limit as $T/P$ grows large, the dynamic terms vanish and extremely eccentric orbits when averaged over time will produce a consistent offset with respect to the planet. Elliptic orbits will thus produce a bias along their major axis. Circular (or near-circular) orbits, as expected, will have a mean position that is centred on the moon-planet barycentre, and so to yield a signal we must observe them over a sufficiently short period that their signal is not averaged out. In practice, this means that it will be difficult to observe moons that require an observation time greater than their period, except for particularly eccentric orbits.

\section{Computing the spectroastrometric signal for closed Keplerian orbits} \label{sec:signal_kepler_derivation}
\noindent To obtain a less unwieldy expression for the spectroastrometric signal, we assume that the observations are taken back-to-back, first the filter $M$ and then the filter $P$. In that case, we obtain for the definition of the spectroastrometric signal (Eq.~\ref{eq:spectroastrometric_signal}) that:
\begin{equation}
    S_{M,P} = \frac{a}{d} \abs{\mathrm{R}_1(i)\mathrm{R}_3(\omega)\left(f_{Mm}\hat{\boldsymbol{\xi}}_M - f_{Pm}\hat{\boldsymbol{\xi}}_P\right)}.
\end{equation}
While this expression looks as though we have not made any progress with respect to Eq.~\ref{eq:spectroastrometric_signal} aside from a rearranging of terms, recall that in the form of Eq.~\ref{eq:xi_app} we have an analytical expression for $\hat{\boldsymbol{\xi}}_M$ and $\hat{\boldsymbol{\xi}}_P$. If we then define $\theta_M$ (respectively $\theta_P$) to be the direction of the vector $\hat{\boldsymbol{\xi}}_M$ (respectively $\hat{\boldsymbol{\xi}}_P$) in the $\xi-\eta$ plane, we find upon a set of applications of the sine and cosine laws in the triangle formed by $f_{Mm}\hat{\boldsymbol{\xi}}_M$, $f_{Pm}\hat{\boldsymbol{\xi}}_P$ and their difference that:
\begin{align}
\label{eq:rotation}
    S_{M,P} = \frac{a}{d}\hat{S}_{M,P} \sqrt{1 - \sin^2{i}\sin^2{\left(\omega  + \theta_M - \alpha\right)}},
\end{align}
where we will call $\hat{S}_{M,P}(e, \tfrac{T_M}{P}, \tfrac{T_P}{P}, f_{Mm}, f_{Pm}, t_0)$  the `dimensionless signal'; the signal if we had observed the orbit face-on in units of $\frac{a}{d}$:
\begin{equation}
    \hat{S}_{M,P} = \abs{ f_{Mm}\hat{\boldsymbol{\xi}}_M - f_{Pm}\hat{\boldsymbol{\xi}}_P}
\end{equation}
and $\alpha$ is the unique angle satisfying simultaneously
\begin{align}
    \sin{\alpha} =& \frac{f_{Pm}\abs{\hat{\boldsymbol{\xi}}_P}}{\hat{S}_{M,P}} \sin{\left(\theta_P-\theta_M\right)} \\
    \cos{\alpha} =& \frac{f_{Mm}\abs{\hat{\boldsymbol{\xi}}_M}}{\hat{S}_{M,P}} - \frac{f_{Pm}\abs{\hat{\boldsymbol{\xi}}_P}\cos{\left(\theta_P - \theta_M\right)}}{\hat{S}_{M,P}}
\end{align}
or equivalently, satisfying the relation:
\begin{equation}
    \alpha = \mathrm{atan2}\left(\sin{\left(\theta_P-\theta_M\right)},\; \frac{f_{Mm}}{f_{Pm}}\frac{\abs{\hat{\boldsymbol{\xi}}_M}}{\abs{\hat{\boldsymbol{\xi}}_P}} - \cos{\left(\theta_P - \theta_M\right)}\right)
\end{equation}
where $\textrm{atan2}(y, x)$ is the 2-argument arctangent.
We note that $\alpha=\alpha(t_0, \tfrac{T_M}{P}, \tfrac{T_P}{P}, \tfrac{f_{Mm}}{f_{Pm}}, e)$, but that it is not a function of $i$ and $\omega$. Hence, all dependency on $i$ and $\omega$ is now expressed explicitly. Moreover, we observe that for $f_{Mm} \gg f_{Pm}$ we have that $\alpha\approx 0$, and that for moons with $e=0$ we have that $\alpha$ is fixed throughout its orbit.

This is nearly in a computable form, except for the fact that we do not know a priori the inclination nor the orientation of the orbit of the moon (as we are trying to establish whether there is a moon in the first place). To ameliorate this, let us take a flat prior on the argument of periapsis, $\omega$, which takes values in $[0, 2\pi]$, and marginalise over this prior. To include circular orbits, we will take $\omega$ to be the reference point on the orbit for $e=0$; this, of course, is equally well-suited to a flat prior and so follows the same mathematical argument. For a face-on orbit (where the line of nodes is degenerate, and so $\omega$ is undefined), the same argument holds if we measure $\omega$ from celestial north instead. This reasoning thus includes all closed orbits.

We justify the flat prior by noting that the argument of periapsis is subject to precession, which for the major satellites of Saturn, for example, is known to occur on timescales of only days to thousands of years \citep{Jacobson2022ThePole}; moreover, theory predicts for the apsidal precession rate a value independent of orientation \citep{Greenberg1981ApsidalPlanet}, such that the argument of periapsis effectively becomes a random variable that is uniformly distributed in time (of course, these timescales are in general not short enough to warrant treating it as such for consequent observations of the same moon). Therefore we can marginalise over $\omega$, and obtain for $S_{M,P}$ (in a small abuse of notation we shall prefer to denote the signal marginalised over $\omega$ and later $i$ as $S_{M,P}$, too);
\begin{align}
    S_{M,P} &= \frac{a}{d} \frac{\hat{S}_{M,P}}{2\pi} \int_0^{2\pi} \sqrt{1 - \sin^2{i}\sin^2{(\omega + \theta_M - \alpha)}} \dd{\omega} \nonumber \\ &= \frac{a}{d} \frac{2\hat{S}_{M,P}}{\pi} E(\sin{i})
\end{align}
where $E(k)=\int_0^{\pi/2}\sqrt{1-k^2\sin^2{x}}\dd{x}$ is the complete elliptic integral of the second kind, which has no solution in terms of elementary functions but has been well-studied and is included in most numerical programming toolkits. If one has knowledge on the inclination of any putative moon, this expression suffices, and there are good arguments to see why one would; formation models predict, for example, that satellites that formed in situ will have negligible inclinations with respect to their host due to inclination damping in the circumplanetary disc (see e.g. \citealt{Moraes2018GrowthDisc}), and most major regular Solar System moons do indeed have near-equatorial inclinations \citep{Musotto2002NumericalSatellites, Jacobson2022ThePole} (though Triton forms a notable exception; see e.g. \citealt{McKinnon2014Triton}); however, out of the Solar System giant planets only Jupiter has an equatorial plane relatively close to the ecliptic \citep{Archinal2018Report2015}, and it is predicted to leave this state over time \citep{Saillenfest2020TheJupiter}. Currently no population-level information about the obliquity of extrasolar planets is known, and recent work has shown that mechanisms exist by which planets may be tilted to obliquities near perpendicular to their orbital angular momentum by migrating moons \citep{Saillenfest2022TiltingSatellite, Saillenfest2023ObliqueRadii}, tilting the plane of their satellite system simultaneously. 

Until such population-level obliquity information becomes available (which could consequently inform moon searches on a per-planet level), then, one will not know the (likely) inclination of the moon and so we must go one step further. Taking a flat prior on the inclination over $i\in[0, \pi]$, we can marginalise over the inclination of the moon, finally arriving at the expected value of the signal for a given semi-major axis and period:
\begin{align}
    S_{M,P} &= \frac{a}{d} \frac{2\hat{S}_{M,P}}{\pi^2} \int_0^{\pi}E(\sin{i}) \dd{i} \\ &= \label{eq:ellipticintegral_version}\frac{a}{d}\hat{S}_{M,P}\left[\frac{2K^2(\sqrt{2}/2)}{\pi^2} + \frac{1}{2K^2(\sqrt{2}/2)}\right] \\&\approx 0.842 \cdot \hat{S}_{M,P}\frac{a}{d}
\end{align}
where $K(k)=\int_0^{\pi/2}\frac{\dd{x}}{\sqrt{1-k^2\sin{x}}}$ is the complete integral of the first kind; the expression for the integral of E in terms of $K(\sqrt{2}/2)$ (that is, Eq.~\ref{eq:ellipticintegral_version}) is due to \citet{Byrd1971HandbookScientists}, Eq.~531.02. This shows that the overall effect of the inclination and orientation of the orbit, while not negligible, is also not too severe. In the best case, with $\sin i = 0$, we have $S_{M,P} = \frac{a}{d} \hat{S}_{M,P}$ (independent of $\omega$), but even in the very worst case for an edge-on orbit with $\sin i = 1$, we find that $S_{M,P} = \frac{2}{\pi}\frac{a}{d}\hat{S}_{M,P}$ such that the signal is reduced to $\sim63.7\%$ its best-case value. We do remark that in this case we have maintained the marginalisation over $\omega$; for a particularly unfortunate realisation of both $i$ and $\omega$ it is indeed possible that the signal at a given time is lower, yet even for $\sin{i}=1$ the moon will spend the greater part of its time in the region where this multiplicative constant is between $\tfrac{2}{\pi}$ and $1$. In the circular case, we note that the dimensionless signal $\hat{S}_{M,P}$ is independent of the time since periapsis, and so the marginalisation over the argument of periapsis effectively amounted to a marginalisation over mean anomaly (i.e. time), too. In that case the expected value in time for the edge-on case is precisely a fraction $2/\pi$ of the face-on value, as a result of the orbital sampling effect previously noted in another context by \citet{Heller2014DetectingEffect}; the moon spends more time at a greater projected distance than it does close to conjunction.

Overall, when observing a random system with no a priori knowledge on the inclination nor orientation of any potential moon, we expect to observe any signal at $\sim 84.2\%$ of its best-case value. For now we shall parametrise this value as $\gamma$, such that $S_{M,P} = \gamma \hat{S}_{M,P}\frac{a}{d}$ with $\gamma\in[\tfrac{2}{\pi}, 1]$ (i.e. of order unity). We then remark that for a circular orbit, the time-dependency of the signal has dropped out during our marginalisation over the argument of periapsis already; for an elliptical orbit, we could marginalise over $t_0\sim U([0, P])$, still, but choose not to, as the resulting integral must be evaluated numerically and does not give any particular insight. The expression that we will use for now thus becomes:
\begin{align}
    S_{M,P} =  \label{eq:calculated_signal_app}\frac{\gamma af_{Mm}}{d}\abs{\hat{\boldsymbol{\xi}}_M - \tfrac{f_{Pm}}{f_{Mm}}\hat{\boldsymbol{\xi}}_P}
\end{align}
where we will explicitly note that $S_{M,P} = S_{M,P}(a, e, \tfrac{T_M}{P}, \tfrac{T_P}{P}, \tfrac{t_0}{P}, f_{Mm}, f_{Pm})$ such that now only the shape of the orbit (in the form of $a$ and $e$), the time at which we observe with respect to last pericentre passage $t_0$ and the spectral characteristics $f_{Mm}$ and $f_{Pm}$ are of interest, and all other quantities are known or have been marginalised out.

\section{Derivation of the noise sources}
Most of the noise sources as summarised in Sec.~\ref{subsec:noise_sources}, to the authors' knowledge, have not yet appeared in previous literature on spectroastrometry, as that has mostly been aimed at spectroastrometry using spectra, not photometry (hence, of course, the name), of high-S/N binary sources (see e.g. \citealt{Bailey1998Spectroastrometry:Scales, Bailey1998DetectionSpectro-astrometry, Porter2004OnFluxes, Whelan2008Spectro-astrometry:Applications}), which do not require such precise noise estimates. Therefore, we feel this warrants a more detailed explanation of their mathematical origin.

\subsection{Derivation of the photon noise}\label{subsec:photon_noise_app}
Though an expression for the photon noise appears in the work by \citet{Agol2015THEEXOMOONS}, we feel that a more rigorous explanation is warranted in the context of the more rigorous derivations we give of the other noise sources. Let us assume we observe the source in a filter over a given time period, in which we observe $N$ photons, of which $f_{m}N$ are due to the moon and $(1-f_{m})N$ due to its host planet (as we will only be discussing a single filter, we will forego the corresponding subscript in the moon flux fraction). For the sake of convenience, let us label the photons $i\in[1, N]$ with the photons $i\in[1, f_mN]$ originating from the moon, and the photons $i\in[f_mN + 1, N]$ originating from the planet. If we then denote the measured location of origin of the photon $i$ as $\mathbf{c}_i$, we can denote total the variance due to the PSF photon noise in the centroid $\sigma_{PN}$ as:
\begin{align}
    \sigma_{PN}^2 &= \mathbb{E}\left[\left(\tfrac{1}{N}\sum_{i=1}^{N}(\mathbf{c}_i - \mathbf{c}_F)\right)^2\right] \nonumber \\
    &= \tfrac{1}{N^2}\sum_{i=1}^{N}\sum_{j=1}^{N}\mathbb{E}\left[(\mathbf{c}_i - \mathbf{c}_F)\cdot(\mathbf{c}_j - \mathbf{c}_F)\right] \label{eq:psf_sum_def}
\end{align}
where $\mathbf{c}_F$ is the centroid position over the measurement, which is given by Eq.~\ref{eq:centroid_location_timeavg}; as we have already solved the effect of the bodies' movement with time in Secs.~\ref{sec:angular_separation_kepler} and \ref{sec:signal_kepler_derivation}, we will neglect this effect for now such that $\mathbf{c}_F=f_m\mathbf{c}_m+(1-f_m)\mathbf{c}_p$. We then observe that we can write $(\mathbf{c}_i - \mathbf{c}_F)$ differently depending on whether the photon $i$ originates from the moon or planet;
\begin{align}
    \mathbf{c}_i - \mathbf{c}_F =
    \begin{cases}
        (\mathbf{c}_i - \mathbf{c}_m) + (1-f_m)\mathbf{c}_{mp} \mbox{ if $i\in[1,f_mN]$} \\
        (\mathbf{c}_i - \mathbf{c}_p) -f_m\mathbf{c}_{mp} \mbox{ if $i\in[f_mN+1, N]$}
    \end{cases}
\end{align}
where we recall that $\mathbf{c}_{mp}=\mathbf{c}_{m} - \mathbf{c}_{p}$. We can therefore divide the summands of Eq.~\ref{eq:psf_sum_def} into several cases: (1) $i=j\in[1,f_mN]$, (2) $i=j\in[f_mN+1, N]$, (3) $i\neq j\in[1,f_mN]$, (4) $i\neq j\in[f_mN+1,N]$, (5) $i\in[1,f_mN]$, $j\in[f_mN+1, N]$ and finally (6) $j\in[1,f_mN]$, $i\in[f_mN+1, N]$. Of course, it must be noted that these cases appear in pairs: (1) and (2), (3) and (4), (5) and (6), the calculation of which is nearly identical. Let us then set for brevity $S(i,j)=\mathbb{E}\left[(\mathbf{c}_i - \mathbf{c}_F)\cdot(\mathbf{c}_j - \mathbf{c}_F)\right]$. If we then (crucially) assume that the PSF of our telescope is point-symmetric such that $\mathbb{E}\left[(\mathbf{c}_i - \mathbf{c}_b)\cdot\mathbf{c}_{mp}\right]=0$ for both $b=m$ and $b=p$, we obtain for each of the six cases the following;
\begin{equation}
    S(i,j) =
    \begin{cases}
        \sigma_{PN}^2 + (1-f_m)^2\mathbf{c}_{mp}^2 & \mbox{ (1)} \\
        \sigma_{PN}^2 + f_m^2\mathbf{c}_{mp}^2 & \mbox{ (2)} \\
        (1-f_m)^2\mathbf{c}_{mp}^2 & \mbox{ (3)} \\
        f_m^2\mathbf{c}_{mp}^2 & \mbox{ (4)} \\
        -f_m(1-f_m)\mathbf{c}_{mp}^2 & \mbox{ (5), (6)} \\
    \end{cases}
\end{equation}
where we have used the fact that $\mathbb{E}\left[(\mathbf{c}_i - \mathbf{c}_b)^2\right]=\sigma_{PSF}^2$ for both $b=m$ and $b=p$, where $\sigma_{PSF}$ is the photon shot noise for a single photon (encapsulated in the PSF distribution), and moreover we have used the fact that any two photons are independent. In Eq.~\ref{eq:psf_sum_def} we note then that there are $f_mN$ copies of case (1), $(1-f_m)N$ copies of case (2), $f_mN(f_mN-1)$ copies of case (3), $(1-f_m)N((1-f_m)N - 1)$ copies of case (4) and a total of $2f_m(1-f_m)N^2$ copies of case (5) and (6) combined. Hence, upon summing as many copies of each term we find that indeed $\sigma_{PN}^2=\tfrac{\sigma_{PSF}^2}{N}$ such that $\sigma_{PN}=\tfrac{\sigma_{PSF}}{\sqrt{N}}$, in agreement with \citet{Agol2015THEEXOMOONS}. What we have gained, however, is the notion that all we need is a point-symmetric PSF, which we already required to show that the measured centroid converges to the geometric centroid in Sec.~\ref{subsec:spectroastrometric_signal}.

\subsection{Derivation of the pixel noise} \label{subsec:pixel_noise_app}
With a robust estimate for the photon noise which accounts for the finite number of photons we count, we are now licensed to discretise the integral in Eq.~\ref{eq:deffiltercentroid} over the photons. Hence, the `actual centroid' for the photons observed over the course of the observation (that is, the average location of origin), which we will call $\mathbf{c}_{ph}$, becomes:
\begin{equation}
    \mathbf{c}_{ph} = \frac{1}{N}\sum_{i=1}^{N}\mathbf{c}_{ph,i} = \frac{1}{N}\sum_{p=1}^{N_{px}}\sum_{i=1}^{N(p)}\mathbf{c}_{ph,i}(p)
\end{equation}
where $\mathbf{c}_{ph,i}$ is the `centroid' (i.e. location of arrival on the detector) of the i-th photon that was detected overall while $\mathbf{c}_{ph,i}(p)$ is the location of arrival of the i-th photon that was detected on a pixel $p$; the reason for this re-indexing will become clear shortly. $N$ is the total number of photons that have struck the detector in this filter, and $N(p)$ is the number of photons that struck pixel $p$. However, with conventional detectors we do not record the location of arrival of the photons; instead, we record that it arrived in a given pixel on our detector. Hence, the total `pixelated' centroid that we compute from the observation data, $\mathbf{c}_{px,tot}$, is given by:
\begin{align}
    \mathbf{c}_{px,tot} = \frac{1}{N} \sum_{p=1}^{N_{px}}\sum_{i=1}^{N(p)}\mathbf{c}_{px}(p)
\end{align}
where $\mathbf{c}_{px}(p)$ is the location of arrival we assign to photons that arrive in the pixel $p$ (we note that this need not be the centre a priori, but we shall see that this follows immediately if we require convergence regardless of the on-sky intensity distribution). We are interested in the `pixel noise': that is, the noise caused by the fact that we assign all photons in the pixel $p$ the location of arrival $\mathbf{c}_{px}(p)$ rather than their actual location of arrival $\mathbf{c}_{ph,i}(p)$, however: we therefore compute the difference of $\mathbf{c}_{ph}$ and $\mathbf{c}_{px,tot}$:
\begin{equation}
\label{eq:pixel_diff}
    \mathbf{c}_{ph} - \mathbf{c}_{px,tot} = \frac{1}{N} \sum_{p=1}^{N_{px}}\sum_{i=1}^{N(p)}\left(\mathbf{c}_{ph,i}(p) - \mathbf{c}_{px}(p)\right).
\end{equation}
It follows upon taking expectations by linearity that $\mathbb{E}\left[\mathbf{c}_{ph} - \mathbf{c}_{px,tot}\right]~=~0$ is satisfied regardless of the intensity distribution when $\mathbb{E}\left[\mathbf{c}_{ph,i}(p) - \mathbf{c}_{px}(p)\right]~=~0$ for all $i$ and $p$ individually. As the on-sky intensity distribution of our source ought not care about the orientation or pixel size of our detector, we should expect that $\mathbf{c}_{ph,i}(p) - \mathbf{c}_{px}(p)$ is uniformly distributed throughout the pixel for each $p$, such that $\mathbb{E}\left[\mathbf{c}_{ph,i}(p) - \mathbf{c}_{px}(p)\right]~=~0$ if and only if $\mathbf{c}_{px}(p)$ is precisely the centre of the pixel, which is therefore a requirement for convergence we must take into account. The pixel noise $\sigma_{px}$ then follows precisely from taking the expectation value of the square of Eq.~\ref{eq:pixel_diff}:
\begin{align}
    \sigma_{px}^2 &= \mathbb{E}\left[\left(\mathbf{c}_{ph} - \mathbf{c}_{px,tot}\right)^2\right] \nonumber \\
    &= \mathbb{E}\left[\left(\frac{1}{N} \sum_{p=1}^{N_{px}}\sum_{i=1}^{N(p)}\left(\mathbf{c}_{ph,i}(p) - \mathbf{c}_{px}(p)\right)\right)^2\right] \nonumber \\
    &= \frac{1}{N^2} \sum_{p=1}^{N_{px}}\sum_{q=1}^{N_{px}}\mathbb{E}\left[S(p)S(q)\right] \label{eq:pixel_sum_exp}
\end{align}
where the last equality follows from expansion of the square of the sum over $p$ into a double-indexed sum over the cross-products of $(p,q)$ (introducing the second index $q$) and we have defined an auxiliary function $S(p)$:
\begin{align}
    S(p) = \sum_{i=1}^{N(p)}\left(\mathbf{c}_{ph,i}(p) - \mathbf{c}_{px}(p)\right).
\end{align}
As we assume any two photons to be statistically independent, any series of photons landing in one pixel must equally well be independent from those landing in another pixel. Hence, $\mathbb{E}[S(p)S(q)]=\mathbb{E}[S(p)]\mathbb{E}[S(q)]$; moreover, $\mathbb{E}[S(p)]=\mathbb{E}[\sum_{i=1}^{N(p)}\left(\mathbf{c}_{ph,i}(p) - \mathbf{c}_{px}(p)\right)]=\sum_{i=1}^{N(p)}\mathbb{E}[\left(\mathbf{c}_{ph,i}(p) - \mathbf{c}_{px}(p)\right)]=0$. As a result, all $(p,q)$ cross-terms in Eq.~\ref{eq:pixel_sum_exp} vanish, and we find:
\begin{align}
    \sigma_{px}^2 &= \frac{1}{N^2} \sum_{p=1}^{N_{px}}\mathbb{E}\left[S(p)^2\right] \nonumber \\
    &= \frac{1}{N^2} \sum_{p=1}^{N_{px}}\mathbb{E}\left[\left(\sum_{i=1}^{N(p)}\left(\mathbf{c}_{ph,i}(p) - \mathbf{c}_{px}(p)\right)\right)^2\right]
\end{align}
where we can again rewrite the square of a sum as an expansion into a double sum over cross-terms. As all photons are independent and $\mathbb{E}\left[\mathbf{c}_{ph,i}(p) - \mathbf{c}_{px}(p)\right]=0$, all cross-terms vanish again and so we are left with
\begin{align}
    \sigma_{px}^2 = \frac{1}{N^2} \sum_{p=1}^{N_{px}}\sum_{i=1}^{N(p)}\mathbb{E}\left[\left(\mathbf{c}_{ph,i}(p) - \mathbf{c}_{px}(p)\right)^2\right]
\end{align}
for which we can produce a worst-case upper bound (that is, a conservative estimate) by noting that regardless of the intensity distribution of the source, $\mathbf{c}_{ph,i}(p) - \mathbf{c}_{px}(p)\leq \alpha/\sqrt{2}$ (where $\alpha$ is the pixel width), as no point in the pixel can be further away from the pixel centre than its vertex, at a distance $\alpha/\sqrt{2}$. A conservative estimate for the pixel noise immediately follows, as then:
\begin{align}
    \sigma_{px}^2 &\approx \frac{1}{N^2} \sum_{p=1}^{N_{px}}\sum_{i=1}^{N(p)}\frac{\alpha^2}{2} = \frac{\alpha^2}{2N}
\end{align}
such that $\sigma_{px}\approx\alpha/\sqrt{2N}$. A less conservative but reasonable estimate can be obtained by assuming that the $\mathbf{c}_{ph,i}$ are uniformly distributed through the pixels (as the intensity distribution should be independent of the orientation and pixel size of our detector); in that case, one obtains $\sigma_{px}=\alpha/\sqrt{6N}$ instead. As the former estimate is guaranteed to be conservative, though, we shall maintain $\sigma_{px}=\alpha/\sqrt{2N}$. Given that, as mentioned, the intensity distribution and our detector properties ought to be independent, it is fair to assume that this effect is independent from the photon shot noise.

\subsection{Derivation of the background and instrument flux noise} \label{subsec:flux_noise_app}
A third effect that will affect the centroid we observe is the inherent random fluctuations in the noise within the area on the detector that we sample. Effectively, this is nothing other than the photon noise term for the noise, but as the noise is evenly distributed over the background sky (or rather, we assume it to be) this term is not contained in our expression for the photon noise from Sec.~\ref{subsec:photon_noise_app}, nor is the derivation fully analogous. We shall therefore derive it separately.

Let us consider the centroid as measured over a noisy observation, $\mathbf{c}_m$, where the noise-corrected observed photon count over each pixel $p$ can be expressed as $N(p)=N_{nf}(p) + N_n(p) - \mu_n$ where $N(p)$ is the photon count due to the source (i.e. planet or moon) in the pixel $p$, $N_{nf}$ is the noise-free photon count in that pixel and $N_n(p)$ is the photon count due to the noise in that same pixel; we will assume that $N_n(p)\sim\textrm{Pois}(\mu_n)$ for all $p$, where $\mu_n$ is the mean noise per pixel. Correspondingly, we shall take $N$, $N_{nf}$ and $N_n$ (without the argument $p$) to be the total number of noise-corrected (i.e. measured), noise-free and noise photons over all pixels that we sample. We will moreover assume that $\mu_n$ has been obtained from an analysis of the source-free parts of the exposure and is therefore known. As we have taken into account the discretisation of our centroid calculation in the photons (through the photon noise) and over the pixels (through the pixel noise) we can now write for the noisy centroid $\mathbf{c}_m$ that
\begin{align}
    \mathbf{c}_m &= \frac{1}{N}\sum_{p=1}^{N_{px}}N_m(p)\mathbf{c}_{px}(p) \nonumber \\
    &= \frac{1}{N}\sum_{p=1}^{N_{px}}(N_{nf}(p) + N_n(p) - \mu_n)\mathbf{c}_{px}(p)
\end{align}
for which we notice that for large $N$ (when $N \gg \abs{N_n - N_{px}\mu_n}$), we have
\begin{align}
    \frac{1}{N}\sum_{p=1}^{N_{px}}N_{nf}(p)\mathbf{c}_{px}(p) \to \frac{1}{N_{nf}}\sum_{p=1}^{N_{px}}N_{nf}(p)\mathbf{c}_{px}(p) = \mathbf{c}_{nf}
\end{align}
with $\mathbf{c}_{nf}$ the noise-free centroid. Hence, we have in that case that
\begin{align}
    \mathbf{c}_m - \mathbf{c}_{nf} = \frac{1}{N}\sum_{p=1}^{N_{px}}(N_n(p) - \mu_n)\mathbf{c}_{px}(p)
\end{align}
such that we obtain (1) by taking expectations and using the fact that $\mathbb{E}\left[N_n(p) - \mu_n\right]=0$ that the measured centroid is unbiased (i.e. still converges to the true centroid) and (2) upon squaring and then taking expectations an expression for the variance due to the background and instrument flux noise, $\sigma_n^2$:
\begin{align}
    \sigma_n^2 = \frac{1}{N^2}\mathbb{E}\left[\left(\sum_{p=1}^{N_{px}}(N_n(p) - \mu_n)\mathbf{c}_{px}(p)\right)^2\right]
\end{align}
whence we have, assuming that the noise fluxes in any two pixels are independent, that
\begin{align}
\label{eq:expectation_term}
    \sigma_n^2 = \frac{1}{N^2}\sum_{p=1}^{N_{px}}\mathbb{E}\left[(N_n(p) - \mu_n)^2\right]\mathbf{c}_{px}(p)^2
\end{align}
but as $N_n(p)\sim\textrm{Pois($\mu_n$)}$, we have that the variance of $N_n(p)$ (which is precisely what the expectation term in Eq.~\ref{eq:expectation_term} is) is equal to $\mu_n$, such that we have
\begin{align}
    \sigma_n^2 = \frac{\mu_n}{N^2}\sum_{p=1}^{N_{px}}\mathbf{c}_{px}(p)^2 \approx \frac{N_n}{N_{px}N^2}\sum_{p=1}^{N_{px}}\mathbf{c}_{px}(p)^2
\end{align}
where we have used that $N_n\approx\mu_n N_{px}$. If we then set the area per pixel $\alpha^2=\Delta\Omega$ and the total area covered by all pixels in our sample region $\Omega=N_{px}\Delta \Omega$, we have
\begin{align}
    \sigma_n^2 = \frac{1}{N}\frac{N_n}{N}\frac{1}{\Omega}\sum_{p=1}^{N_{px}}\Delta\Omega\mathbf{c}_{px}(p)^2.
\end{align}
We then recognise (1) the ratio $N/N_n$ as the signal-to-noise ratio of the planet detection in the exposure, which we shall set to 5 as a lower bound, and (2) the sum as the (polar) second moment of area of the pixel region over which we sample with respect to the origin of our coordinate system $J~=~\int\int\mathbf{c}^2\dd{\Omega}$; we should therefore set our coordinate system at the centre of the region over which we sample. Moreover, this region ought to minimise the quantity $J/\Omega$; hence, there is clearly a trade-off between sampling a sufficiently large region so as to achieve the greatest photon count from the planet and moon on the one hand, and minimising the ratio $J/\Omega$ on the other hand. A full numerical analysis could potentially yield an optimal area, but we shall simply take at least the $3\sigma$-enclosing region for the PSF (which contains $>97\%$ of the flux). As we have a pixelated region, we cannot achieve the optimal shape for the lowest $J/\Omega$ (which would be a circle), but we can give an `upper bound' or worst-case shape that will always include the $3\sigma$-enclosing region, which is then a square of sidelengths $\lceil6\sigma_{PSF}/\alpha\rceil\alpha$ (the ceiling function is necessary, as we can only sample in full pixels), such that we have an upper bound for $J/\Omega=\lceil6\sigma_{PSF}/\alpha\rceil^2\alpha^2/6$. Hence, we have
\begin{align}
    \sigma_n^2 \leq \frac{1}{30N}\left\lceil\frac{6\sigma_{PSF}}{\alpha}\right\rceil^2\alpha^2
\end{align}
whence
\begin{align}
    \sigma_n \leq \sqrt{\frac{6}{5}}\left(\frac{1}{6}\left\lceil\frac{6\sigma_{PSF}}{\alpha}\right\rceil\right)\frac{\alpha}{\sqrt{N}}
\end{align}
which we shall therefore adopt as a conservative upper bound. We note that while this is an upper bound, for a complete measurement one can in practice calculate an exact estimate for this term a posteriori (as then the signal-to-noise ratio of the detection in the flux and the sampling region are known exactly).
\end{appendix}
\end{document}